\documentclass[11pt,letterpaper]{article}
 \pdfoutput=1

\usepackage{jcappub}

\usepackage{caption}
\usepackage{floatrow}
\usepackage{journals}

\usepackage{subfig}

\usepackage{bm}

\def\be{\begin{eqnarray}}
\def\ee{\end{eqnarray}}

\def\bx{\mathrm{\textbf{COLA}}}

\def\x{\bm{x}}

\def\q{\bm{q}}
\def\s{\bm{s}}

\def\v{\bm{v}}

\usepackage{array}
\newcolumntype{L}[1]{>{\raggedright\let\newline\\\arraybackslash\hspace{0pt}}m{#1}}
\newcolumntype{C}[1]{>{\centering\let\newline\\\arraybackslash\hspace{0pt}}m{#1}}
\newcolumntype{R}[1]{>{\raggedleft\let\newline\\\arraybackslash\hspace{0pt}}m{#1}}

\newsavebox{\tempbox}

\title{sCOLA: The N-body COLA Method Extended to the Spatial Domain}
\author{Svetlin Tassev$^{a,b}$, Daniel J. Eisenstein$^{b}$, Benjamin D. Wandelt$^{c,d,e,f}$, Matias Zaldarriaga$^{g}$}
\affiliation{ 
\sl $^{a}$ Braintree High School, 128 Town Street, Braintree, MA 02184, USA\\
\sl $^{b}$ Center for Astrophysics, Harvard University, 60 Garden Street, Cambridge, \\MA 02138, USA\\
\sl $^{c}$ Institut d'Astrophysique de Paris (IAP), UMR 7095, CNRS - UPMC Universit\'e Paris 6, 98bis boulevard Arago, F-75014 Paris, France\\
\sl $^{d}$ Institut Lagrange de Paris (ILP), Sorbonne Universit\'es, 98bis boulevard Arago, F-75014 Paris, France\\
\sl $^{e}$ Department of Physics, University of Illinois at Urbana-Champaign, Urbana, \\IL~61801, USA\\
\sl $^{f}$ Department of Astronomy, University of Illinois at Urbana-Champaign, Urbana, \\IL~61801, USA\\
\sl $^{g}$ School of Natural Sciences, Institute for Advanced Study, Olden Lane, Princeton, \\NJ 08540, USA}

\emailAdd{stassev@alum.mit.edu}

\abstract{We present sCOLA -- an extension of the N-body COmoving Lagrangian Acceleration (COLA) method to the spatial domain. Similar to the original temporal-domain COLA, sCOLA is an N-body method for solving for large-scale structure in a frame that is comoving with observers following trajectories calculated in Lagrangian Perturbation Theory. Incorporating the sCOLA method in an N-body code allows one to gain computational speed by capturing the gravitational potential from the far field using perturbative techniques, while letting the N-body code solve only for the near field. The far and near fields are completely decoupled, effectively localizing gravity for the N-body side of the code. Thus, running an N-body code for a small simulation volume using sCOLA can reproduce the results of a standard N-body run for the same small volume embedded inside a much larger simulation. We demonstrate that sCOLA can be safely combined with the original temporal-domain COLA. sCOLA can be used as a method for performing zoom-in simulations. It also allows N-body codes to be made embarrassingly parallel, thus allowing for efficiently tiling a volume of interest using grid computing. Moreover, sCOLA can be useful for cheaply generating large ensembles of accurate mock halo catalogs required to study galaxy clustering. Surveys that will benefit the most are ones with large aspect ratios, such as pencil-beam surveys, where sCOLA can easily capture the effects of large-scale transverse modes without the need to substantially increase the simulated volume. As an illustration of the method, we present proof-of-concept zoom-in simulations using a freely available sCOLA-based N-body code.
}

\notoc
\usepackage{graphicx}
\begin{document}
\maketitle

\section{Introduction}

Understanding Large Scale Structure (LSS) of the universe   has been a multi-generational effort in understanding the macroscopic physics governing the evolution of the universe, arising from a trivial microscopic law -- the law of universal gravitation (corrected for the expansion of the universe, as well as baryonic physics at small scales). Analytical approaches to the problem have recently (e.g. \cite{2012JHEP...09..082C,2013MNRAS.429.1674C,2014JCAP...05..022P}) produced new insights into the physics of LSS, significantly improving the predictive power of analytical techniques. Yet, numerical simulations are indispensable when one needs to go beyond the mildly non-linear regime, as self-consistent analytical approaches break down beyond the non-linear scale, corresponding to a wavevector of $\sim0.5$Mpc$/h$ at redshift of zero.

As the volume and precision of observational surveys of LSS increase, modelling LSS observables  (e.g. galaxies, weak lensing) has been pushing the limits of both analytical and numerical techniques. One of the big challenges is accurately modelling uncertainties. That requires reducing sample variance for the 4-point function (covariance matrices) of the tracer field of LSS of choice for a particular survey. The problem can in principle be solved by running hundreds and even thousands of numerical simulations, but until recently this was feasible only after introducing rather crude approximations (e.g. the PTHalos method \cite{2002MNRAS.329..629S, 2012arXiv1203.6609M}). 

The push to construct crude, yet realistic, numerical simulations to tackle the problem of modeling covariances has led to the development of numerous prescriptions \cite{2013JCAP...06..036T,2013JCAP...11..048L,2013MNRAS.433.2389M,2014MNRAS.437.2594W,2014arXiv1407.1236K,2014arXiv1412.5228A,2015MNRAS.446.2621C} for fast generation of density fields and/or mock halo and galaxy catalogs (for a recent review, see \cite{2014arXiv1412.7729C}). One of them, the COmoving Lagrangian Acceleration (COLA) method  (\cite{2013JCAP...06..036T}, hereafter referred to as TZE), is a proposed modification to existing N-body codes, which allows for the reuse of post-processing pipelines, designed to work with N-body codes. COLA allows for a fine control over the trade-off between accuracy and speed and as a cheap N-body method it has been successfully applied in extracting cosmological information from observations \cite{2014MNRAS.441.3524K,2014arXiv1409.3238H,2014arXiv1409.3242R,2014arXiv1410.0355L,2015arXiv150202690L}. 

In its original formulation, the COLA method used the fact that the temporal evolution of large-scale modes is well described by Lagrangian Perturbation Theory (LPT) \cite{1995A&A...296..575B} (see \cite{2014JCAP...06..008T} for a recent detailed discussion of LPT). COLA decouples the temporal evolution of the large and small scales in N-body codes by evolving the large scales using the exact LPT results for the growth factor of those scales, while letting the N-body code calculate the time evolution of the small scales.

In its approach of combining Perturbation Theory (PT) with N-body simulations,  COLA is taking advantage only of the fact that the temporal behavior of large scales can be modeled by LPT. Yet, there is still room for improvement, as PT can also be used to describe how distant objects gravitatonally influence local dynamics. Such far-field effects are dominated by large scales -- most importantly by rescaling the local matter density and introducing tidal effects. The reason for that is that the detailed small-scale distribution of matter far away from a region of interest only changes the higher multipoles (in a multipole expansion) of the potential which decay much faster than the low multipoles (which are affected by the large scales). 

A method that allows for the complete decoupling of local dynamics (modeled with an N-body) from far-field effects (modeled with PT) can have numerous applications. An obvious one is performing zoom-in simulations without the need to substantially extend the simulation volume around a region of interest. It would also allow N-body codes to be made embarrassingly parallel, thus allowing for efficiently tiling a volume of interest. Moreover, such a method can be useful for cheaply generating large ensembles of accurate mock halo catalogs required to study galaxy clustering. Surveys that would benefit the most are ones with large aspect ratios, such as pencil-beam surveys, where such a method can capture the effects of large-scale transverse modes in PT without the need to substantially increase the simulated volume\footnote{An approach to solve this problem has been put forward by \cite{2010ApJS..190..311C}, which uses the properties of the periodic boundary conditions to map a simulation done on a periodic cube into elongated box-like shapes. However, this approach breaks the behavior of the large modes by changing the topological connectedness of the simulated region.}. 

In this proof-of-concept paper we propose exactly such a modification to existing N-body codes by extending the COLA method to the spatial domain, resulting in what we call \textit{spatial} COLA (sCOLA). In Section~\ref{sec:general} we present the general idea behind the sCOLA method. We present results from a set of zoom-in sCOLA-based N-body simulations in Section~\ref{sec:ex}, and we summarize in Section~\ref{sec:sum}. We leave the detailed description of the sCOLA method to the appendices.

\section{Augmenting Simulations with Perturbation Theory to Model the Far Field}\label{sec:general}

\subsection{Review of the original temporal COLA method}

Since sCOLA is an extension of COLA, we start with a quick review of the original COLA method. In COLA (see TZE) one starts with the standard equation of motion for cold dark matter (CDM), which schematically states
\be\label{sketchEoM}
\partial_t^2\x=-\bm{\nabla} \nabla^{-2}\delta[\x](\x)\ .
\ee
In COLA, one rewrites the above equation as
\be\label{sketchEoMLPT}
\partial_t^2\x_{\mathrm{res}}=-\bm{\nabla} \nabla^{-2}\delta[\x](\x)-\partial_t^2\x_{\mathrm{LPT}}\ , \ \ \hbox{with} \ \x_{\mathrm{res}}\equiv \x-\x_{\mathrm{LPT}}\ .
\ee
In the above, $\bm{\nabla}\nabla^{-2}\delta[\bm{a}](\bm{b})$ is a short-hand for the gradient of the inverse Laplacian of the fractional CDM overdensity arising from a distribution of particles at positions $\bm{a}$, evaluated at position $\bm{b}$ of the particle for which the equation of motion is solved for. Here $\x(\q,t)=\x_{\mathrm{LPT}}+\x_{\mathrm{res}}$   is the Eulerian position of CDM particles at time $t$, which started out at (Lagrangian) position $\q$; $\x_{\mathrm{LPT}}$ is the LPT approximation to $\x$; and $\x_{\mathrm{res}}$ is the (residual) displacement of CDM particles as measured in a frame comoving with ``LPT observers'', whose trajectories are given by $\x_{\mathrm{LPT}}$. Note that $\x_{\mathrm{LPT}}(\q,t)$ is the perturbative solution to (\ref{sketchEoM}):
\be\label{lpteq}
\partial_t^2\x_{\mathrm{LPT}}= -\bm{\nabla} \nabla^{-2}\delta[\x_{\mathrm{LPT}}](\x_{\mathrm{LPT}}) \ \ \  \hbox{(solved in PT)} ,
\ee
where the equation above is solved using the standard LPT prescription of expanding it in powers of the linear overdensity.

In COLA, one discretizes the $\partial_t^2$ operator on the left-hand side of (\ref{sketchEoMLPT}), while calculating $\partial_t^2\x_{\mathrm{LPT}}$ on the right-hand side exactly in LPT. Therefore, for zero timesteps, COLA recovers the LPT results, which implies that large scales are followed with an accuracy of at least that of LPT. Corrections to LPT are obtained by increasing the number of timesteps -- a characteristic especially important for recovering the correct time evolution of small scales. When the number of timesteps becomes large, COLA reduces to a standard N-body code. In the intermediate regime, (for $\mathcal{O}(10)$ timesteps) COLA is able to give a good approximation of halo statistics at the expense of not resolving the detailed particle trajectories inside halos (see TZE and  \cite{2014arXiv1412.7729C} for further discussion).

\subsection{Introducing sCOLA}

As discussed in the introduction, the sCOLA method is a modification to existing N-body codes allowing one to perform simulations without the need to substantially extend the simulated volume beyond the region of interest in order to capture far-field effects, such as the effects of super-box modes (modes longer than the size of the simulation box). To understand the principles behind sCOLA, we will first state the equations behind sCOLA and will only then proceed to discuss their interpretation and characteristics in detail.

Let us imagine that we want to simulate the detailed dynamics of a region of interest (referred to as the COLA volume/box from now on), which is embedded in a much larger region (the ``full'' volume/box). sCOLA stipulates that one can then model the dynamics inside the COLA volume by rewriting the equation of motion (\ref{sketchEoM}) as (we reintroduce the omitted constants and Hubble expansion in Appendix~\ref{app:KDK}):
\be\label{sCOLA1}
\partial_t^2\x_{\mathrm{res}}=-\bm{\nabla} \nabla^{-2}_\bx\delta[\x](\x)-\partial_t^2\x^{\bx}_{\mathrm{LPT}}\ , \ \ \hbox{with} \ \x_{\mathrm{res}}\equiv \x-\x_{\mathrm{LPT}}\ ,
\ee
where we introduced an auxiliary LPT displacement field, $\x^{\bx}_{\mathrm{LPT}}(\q,t)$, which similar to (\ref{lpteq}) is defined as the perturbative solution to  (we reintroduce the omitted constants and Hubble expansion in Appendix~\ref{app:KDKlpt})
\be\label{sCOLA2}
\partial_t^2\x^{\bx}_{\mathrm{LPT}}= -\bm{\nabla} \nabla^{-2}_\bx\delta[\x_{\mathrm{LPT}}](\x_{\mathrm{LPT}}) \ \ \  \hbox{(solved in PT)} .
\ee
The subscript \textbf{COLA} in $\nabla^{-2}_\bx$ stands for the fact that the inverse Laplacian used to obtain the gravitational potential from the density field is restricted only over the near field (the COLA volume). In other words, the potential computed with $\nabla^{-2}_\bx\delta$ includes only particles in the COLA volume, not the full volume. Compare that to $\nabla^{-2}$ in (\ref{sketchEoMLPT}), which is over the full box. This marks the crucial difference between the original temporal domain COLA and sCOLA. Equations (\ref{sCOLA1}) and (\ref{sCOLA2}) solved for $\x$ for the particles inside the COLA volume summarize the sCOLA method.

One can think of the sCOLA equations in the following way. We calculate the acceleration of each particle in the COLA box in the frame of an accelerated ``LPT observer'', who follows a  trajectory ($\x_{\mathrm{LPT}}^\bx$) as specified by LPT in that same COLA volume. We then use that acceleration to calculate the extra displacement ($\x_{\mathrm{res}}$) of the particle, which we need to add to the LPT displacement in order to find the actual position of the particle. However, instead of adding $\x_{\mathrm{res}}$ to $\x_{\mathrm{LPT}}^\bx$, which is calculated in the COLA box, we add $\x_{\mathrm{res}}$ to the LPT position ($\x_{\mathrm{LPT}}$) as calculated in the full box. 

Simply put, the idea is that by using the wrong observer ($\x_{\mathrm{LPT}}^\bx$) in the wrong universe (by calculating accelerations in the COLA box) and then pretending we did no such thing (by adding the true LPT position, $\x_{\mathrm{LPT}}$), we make two mistakes, which will tend to cancel out. We demonstrate that that is indeed what happens in the rest of this paper.

\subsection{Discussion}\label{scola:disc}

There are several important aspects of the sCOLA equations that we highlight below\footnote{Note that sCOLA is somewhat similar to existing post-processing methods for capturing the effects on N-body simulations from changes in the background cosmology \cite{2010MNRAS.405..143A} or for resampling the large-scale modes of  existing simulations \cite{2011ApJ...737...11S}. However, unlike sCOLA, both of those methods fail to capture the coupling between the large- and short-scale displacement fields and we will, therefore, not consider them further.}:

1) Once the initial conditions $\x_{\mathrm{LPT}}$ are known, the sCOLA equations above make no reference to particles or fields (be that density, potential or otherwise) outside of the COLA volume. Thus, the non-locality of gravity arising from the inverse Laplacian in the calculation of the gravitational potential, has been severely limited to the COLA volume. Therefore, a sCOLA-based N-body code solving for the positions of particles inside the COLA volume need only work with particles and fields inside of that small volume and not the full volume, which the COLA volume is embedded in. Thus, the main objective of sCOLA -- to decouple the near and far fields, using PT for the far field, and an N-body for the near field -- has been achieved. 

2) As mentioned above, the only place where the far field enters is when one calculates the initial conditions $\x_{\mathrm{LPT}}$ in the full box, as the LPT displacements of the particles in the COLA box are correlated with  the LPT displacements of all particles in the full box (albeit to different degrees). One should generate those initial conditions in the same way one would for a standard N-body run. For zoom-in simulation, for example, one can use multi-scale initial conditions generators such as those presented in  \cite{music} or \cite{panphasia}. 

3) The sCOLA equations become exact in the limit when the COLA volume matches the full volume, reducing to the original COLA equation, eq.~(\ref{sketchEoMLPT}). When those two volumes do not coincide, the error one makes is given by the residual acceleration defined by:
\be
\bm{a}_{\mathrm{res}}(\q)&\equiv& \left.\bigg(\bm{\nabla}\nabla^{-2}_\bx \delta[\x_{\mathrm{LPT}}](\x_{\mathrm{LPT}}) -\bm{\nabla}\nabla^{-2} \delta[\x_{\mathrm{LPT}}](\x_{\mathrm{LPT}})\bigg)\right|_{\hbox{in PT}} \nonumber \\
&-&\bigg(\bm{\nabla}\nabla^{-2}_\bx \delta[\x](\x) -\bm{\nabla}\nabla^{-2} \delta[\x](\x)\bigg) \ ,
\ee
where the first term is calculated in PT, and the right-hand side is evaluated for the particle/LPT observer starting at $\q$. 
Clearly, if the COLA volume matches the full volume, then the inverse Laplacians coincide, $\bm{a}_{\mathrm{res}}$ vanishes,  and sCOLA reduces to the original COLA method as mentioned above. When the COLA volume does not cover the full volume, then $\bm{a}_{\mathrm{res}}$ represents corrections coming only from the far field (the field not covered by the COLA volume). As LPT is an extremely good approximation to the density field at large scales (e.g. \cite{2014JCAP...06..008T}), then $\bm{a}_{\mathrm{res}}$ inside the COLA volume contains \textit{only} corrections from the far field that come \textit{only} from small-scale deviations of the density field from its LPT prediction. Thus, as long as the COLA box is not too small (see below), then $\bm{a}_{\mathrm{res}}$ will contain only the gradients of  quickly decaying high-order multipoles of the far field.

4) From the discussion in (3) above, we see that the errors arising from $\bm{a}_{\mathrm{res}}$ are most prominent near the boundaries of the COLA volume. Thus, one should always make sure to pad their region of interest by roughly the scale below which LPT breaks down (see also (3) above), which is about 10$\,$Mpc$/h$ at redshift zero (e.g. \cite{2012JCAP...12..011T}). One should also make sure that their COLA volume encloses all particles that eventually make it into their region of interest. The typical distance CDM particles travel to redshift zero is again about 10$\,$Mpc$/h$   (e.g. \cite{2014JCAP...06..008T}). Thus, as a rule of thumb, for a simulation down to redshift of zero, the COLA volume should cover a region which is roughly $\sim$10$\,$Mpc$/h$ larger in all directions around the region of interest. The actual number will depend on the specific application of sCOLA and the desired accuracy.

5) One may wonder what boundary conditions (BCs) one should use for the COLA volume. The short answer is: any will probably do as long as those BCs are applied consistently in both sCOLA equations. For any reasonable COLA volume (see above), errors arising from the BCs will only modify the amplitudes of the far-field high-order multipoles entering $\bm{a}_{\mathrm{res}}$, which, as we already discussed, decay rapidly away from the boundaries of the COLA volume.

6) The time operator in the sCOLA equations can be discretized either in the standard approach (discretize all $\partial_t^2$), or in the original temporal domain COLA approach, where one discretizes $\partial_t^2$ only on the left-hand side of (\ref{sCOLA1}) as the time evolution of $\x_{\mathrm{LPT}}$ and $\x_{\mathrm{LPT}}^\bx$ is known to be governed by the usual linear (and second order, for second order LPT) growth factor. In this paper we use the latter approach which will allow us to demonstrate that the temporal and spatial versions of COLA can be used successfully both separately and in conjunction.

\begin{table}[t!]
\caption{The table summarizes the properties of the central halo as obtained from each snapshot (identified by $n_t$ and L$_{\mathrm{COLA}}$) shown in the figures. The halo properties shown are the halo mass, the  center-of-mass position, the halo size  (defined as the root-mean-square halo particle positions); the  center-of-mass halo  velocity, scaled to show the shift of the halo position in redshift space; and the velocity dispersion (defined as the rms halo particle velocity). Apart from the reference results, all other results for the halo properties show the error with respect to the reference result. The errors are absolute unless shown as percentage difference. 
 Reducing the number of timesteps for COLA naturally degrades the quality of the simulation. However, note that the sCOLA-side of the method proves robust to reducing the size of the COLA box until the COLA boundary region (at redshift zero, corresponding to the region $\sim10$\,Mpc$/h$ or less from the edge) invades the region of interest.} \label{table:nt}
\begin{tabular}{|C{1.3cm}|C{1.3cm}|C{1.7cm}|C{1.9cm}|C{1.3cm}|C{1.9cm}|C{1.9cm}|}
\hline 
L$_{\mathrm{COLA}}$ [Mpc$/h$]  & $n_t$ & Mass [M$_\odot/h$]& Mean Position [Mpc$/h$] & Size [Mpc$/h$]   & Mean Velocity  [Mpc$/h$] & Velocity Dispersion [Mpc$/h$] \\ 
\hline
\hline
\multicolumn{7}{ |c| }{\normalsize Reference results ($L=$L$_\mathrm{COLA}$, $n_t=100$)} \\
\hline 
100 &100&  $2.08\times10^{14}$& 0 &  0.79   &  3.18 & 9.65 \\
\hline \hline
\multicolumn{7}{ |c| }{\normalsize Original temporal COLA results relative to reference} \\
\multicolumn{7}{ |c| }{$L=$L$_\mathrm{COLA}$} \\
\hline 
100 & 15 &  -15 \%& 0.10 & +15 \%     & 0.23 &  -17 \% \\
\hline 
100 & 10 &  -18 \% & 0.17 & +25 \%    & 0.30 &  -21 \% \\
\hline 
100 & 7  &  -38 \% &  0.06& +31 \%    & 0.43 &  -30 \% \\
\hline 
\hline 
\multicolumn{7}{ |c| }{\normalsize sCOLA results relative to reference} \\
\multicolumn{7}{ |c| }{$n_t=100$} \\
\hline
75   &  100 &  -1    \%& 0.01&  +1  \%    & 0.06  &  -0.3  \%\\
\hline
50   &  100 &  -0.1  \% & 0.03&  +2  \%  & 0.04  &  -0.4  \%\\
\hline
25  &  100 &  -2    \%& 0.25 &  +68 \%    & 0.37  &  -11   \%\\
\hline \hline
\multicolumn{7}{ |c| }{\normalsize sCOLA results relative to reference} \\
\multicolumn{7}{ |c| }{$n_t=10$} \\
\hline
100 &   10 &   -18 \%&  0.17 &  +25 \%  &  0.30 &  -21 \%\\
\hline
75  &     10 &  -19 \%&  0.18&  +26 \%  &  0.30&  -22 \%\\
\hline
50  &   10 &   -19 \%&  0.18&  +26 \%  &  0.25 &  -22 \%\\
\hline
25   &   10 &   -20 \% &  0.50&  +30 \% & 0.20 &  -24 \%\\
\hline 
\end{tabular} 
\end{table}

\section{Illustrative example}\label{sec:ex}

To illustrate how sCOLA can be implemented in an N-body code, we developed a freely available and open-source parallel Python\footnote{\url{https://www.python.org/}} code, implementing in conjunction the sCOLA and the original COLA methods, dubbed pyCOLA.  To generate the initial conditions for the simulations presented below, we use the multi-scale initial conditions generator, MUSIC\footnote{Note that while working on this paper, we encountered a bug in MUSIC, which resulted in a wrong second-order displacement field. O. Hahn has kindly found and fixed the bug in version 1.51 of MUSIC.} \cite{music}. pyCOLA then calculates the LPT displacements in the COLA volume, 
$\x_{\mathrm{LPT}}^\bx$, by implementing the equations in Appendix~\ref{app:KDKlpt}, corresponding to the schematic eq.~(\ref{sCOLA2}). pyCOLA then solves for  the particle trajectories in the COLA box by evolving the particle positions and velocities using the kick-drift-kick method in Appendix~\ref{app:modKDK}, corresponding to the schematic eq.~(\ref{sCOLA1}). pyCOLA implements periodic BCs for the COLA volume and the force calculation is done using a PM grid over the same COLA volume. We only include particles that are present in the COLA volume in the initial conditions, and do not include particles that eventually enter that region from the outside.

\begin{figure}[t!]
\centering
\subfloat{\includegraphics[width=0.248\textwidth]{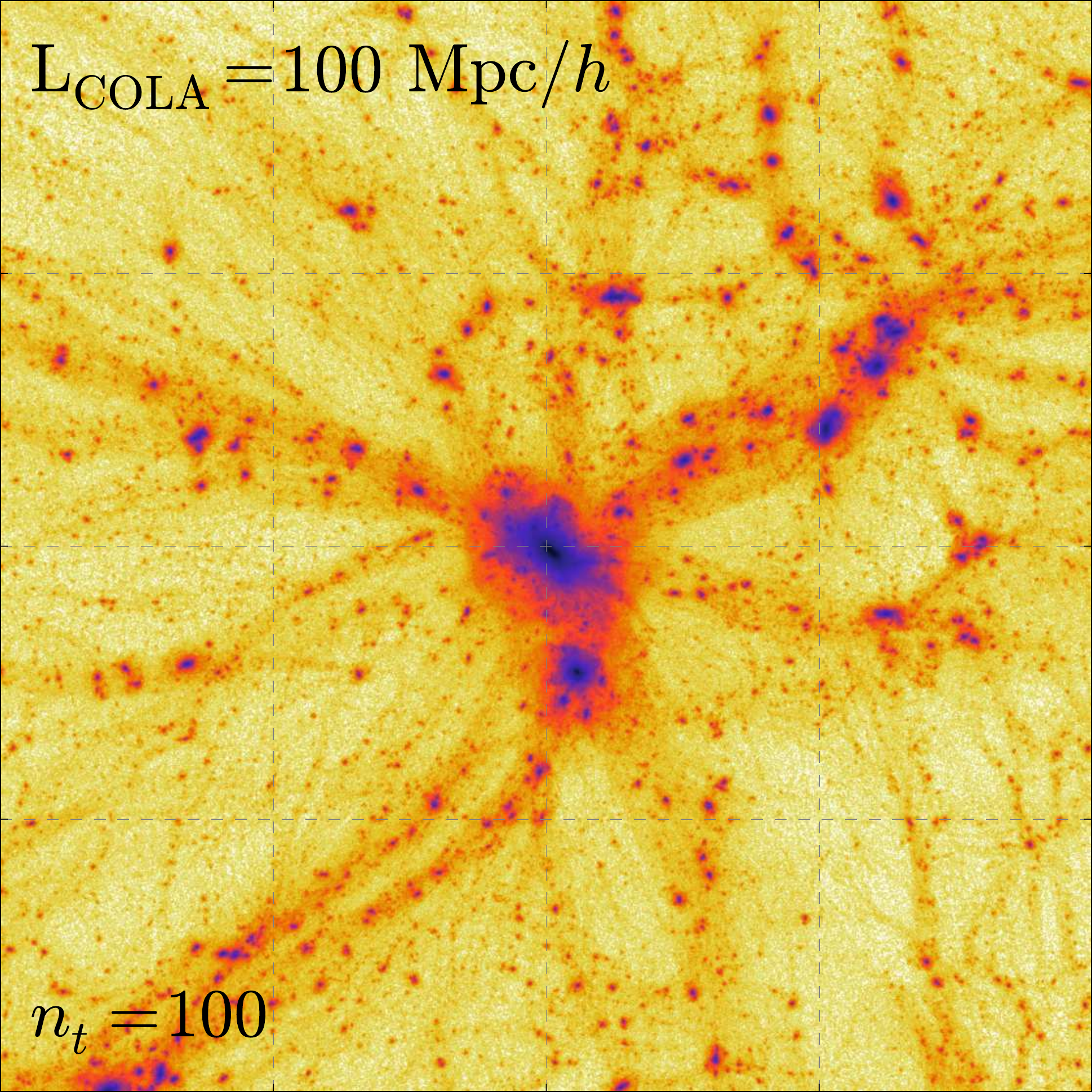}}
\subfloat{\includegraphics[width=0.248\textwidth]{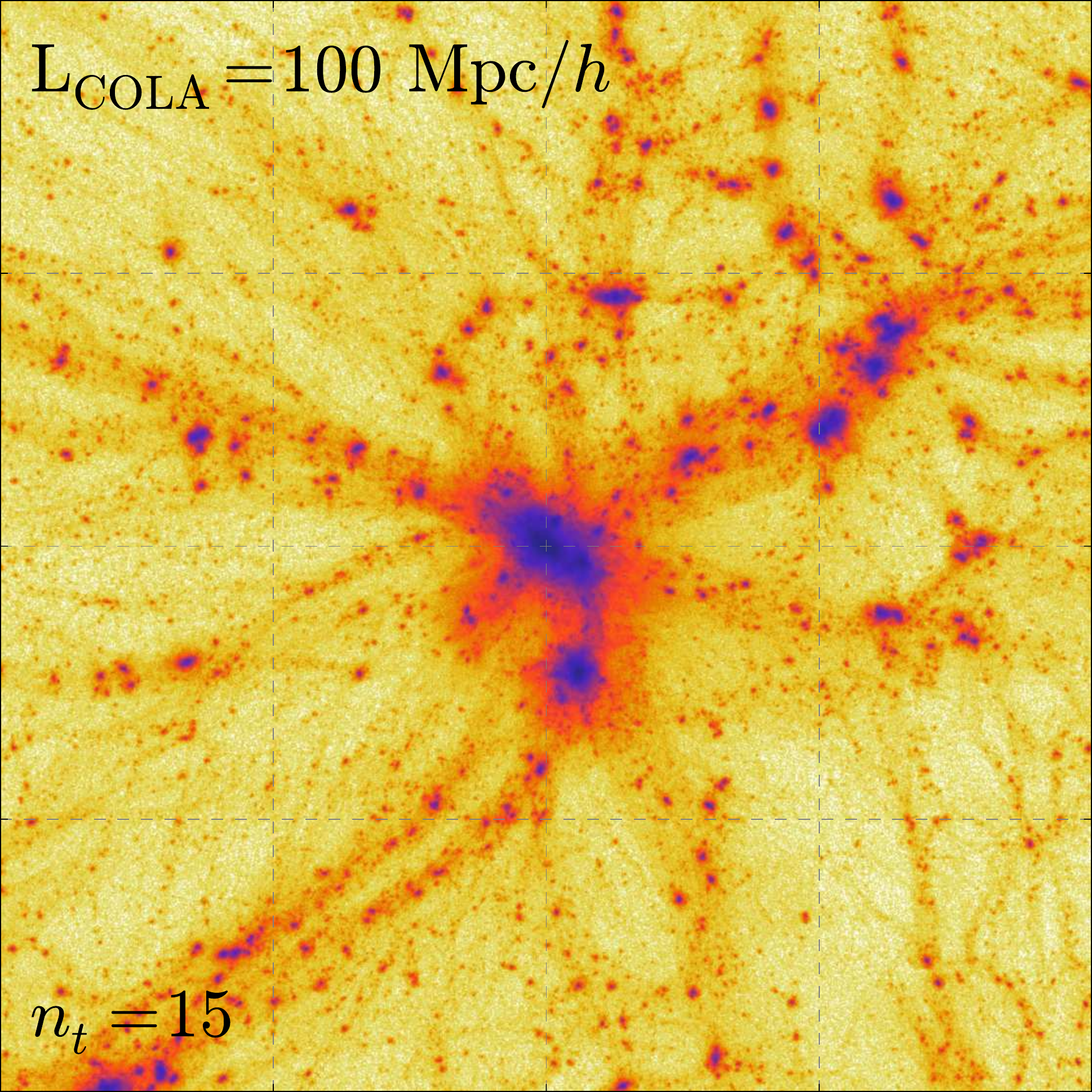}}
\subfloat{\includegraphics[width=0.248\textwidth]{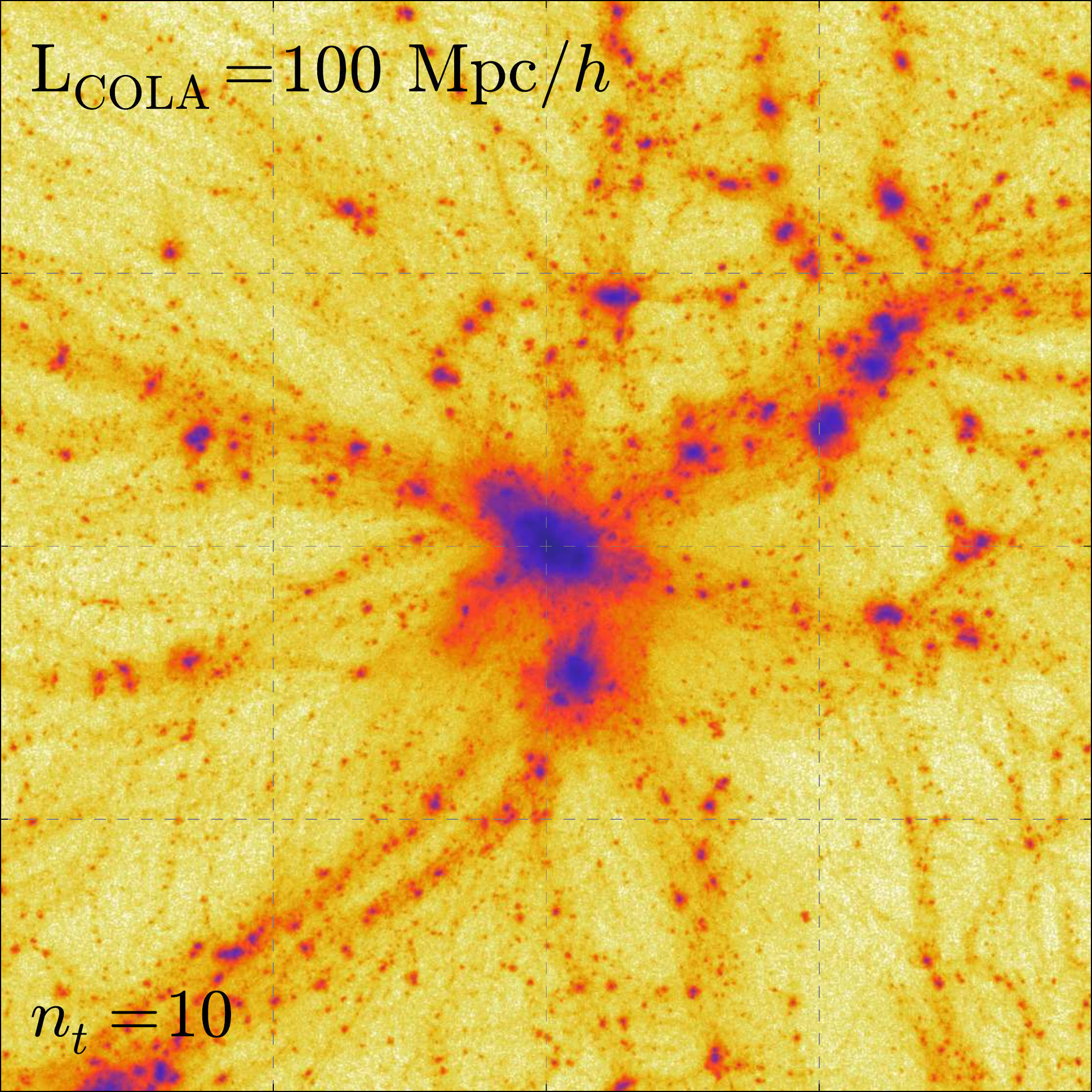}}
\subfloat{\includegraphics[width=0.248\textwidth]{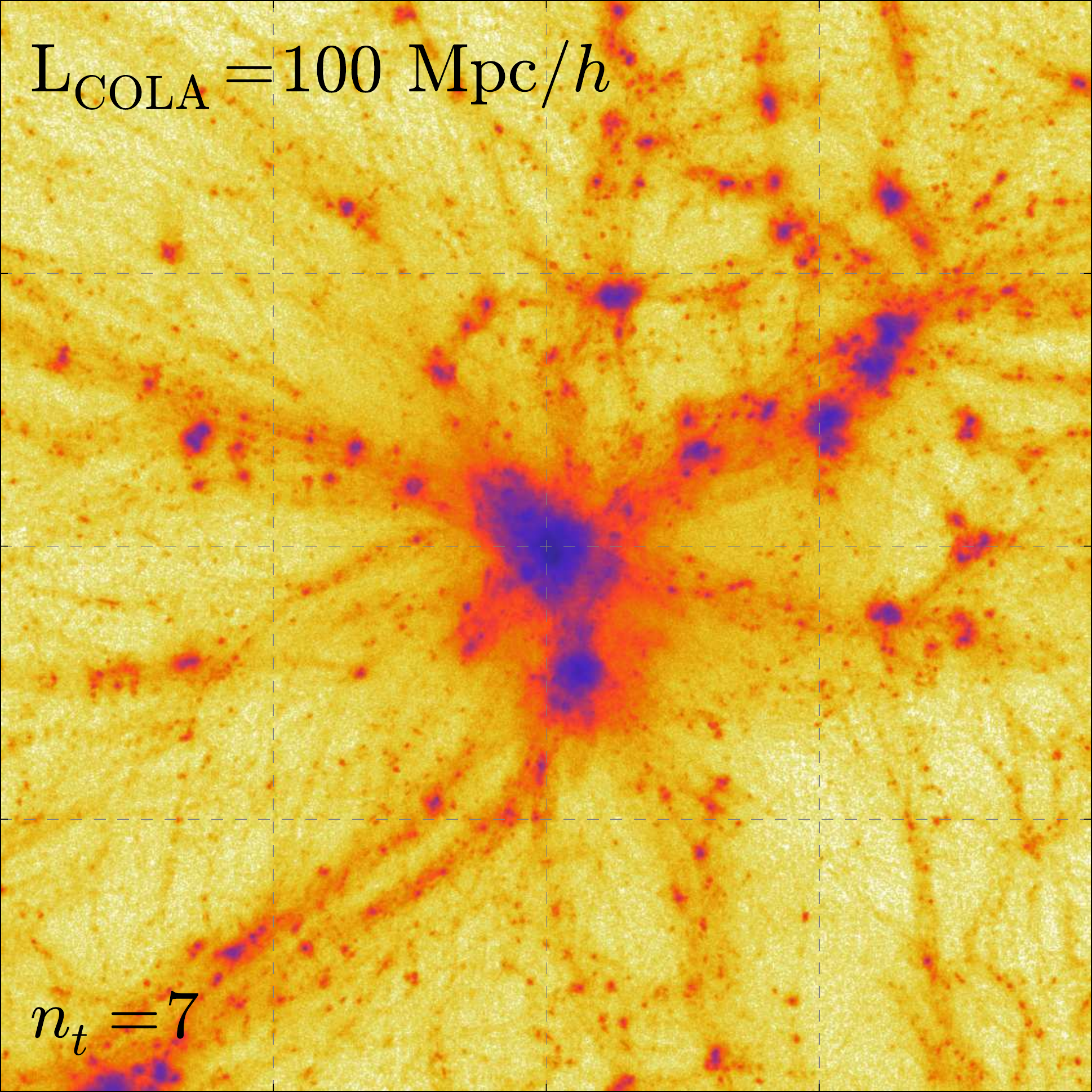}}
\hfill
\\
\vspace{-0.38cm}
\subfloat{\includegraphics[width=0.248\textwidth]{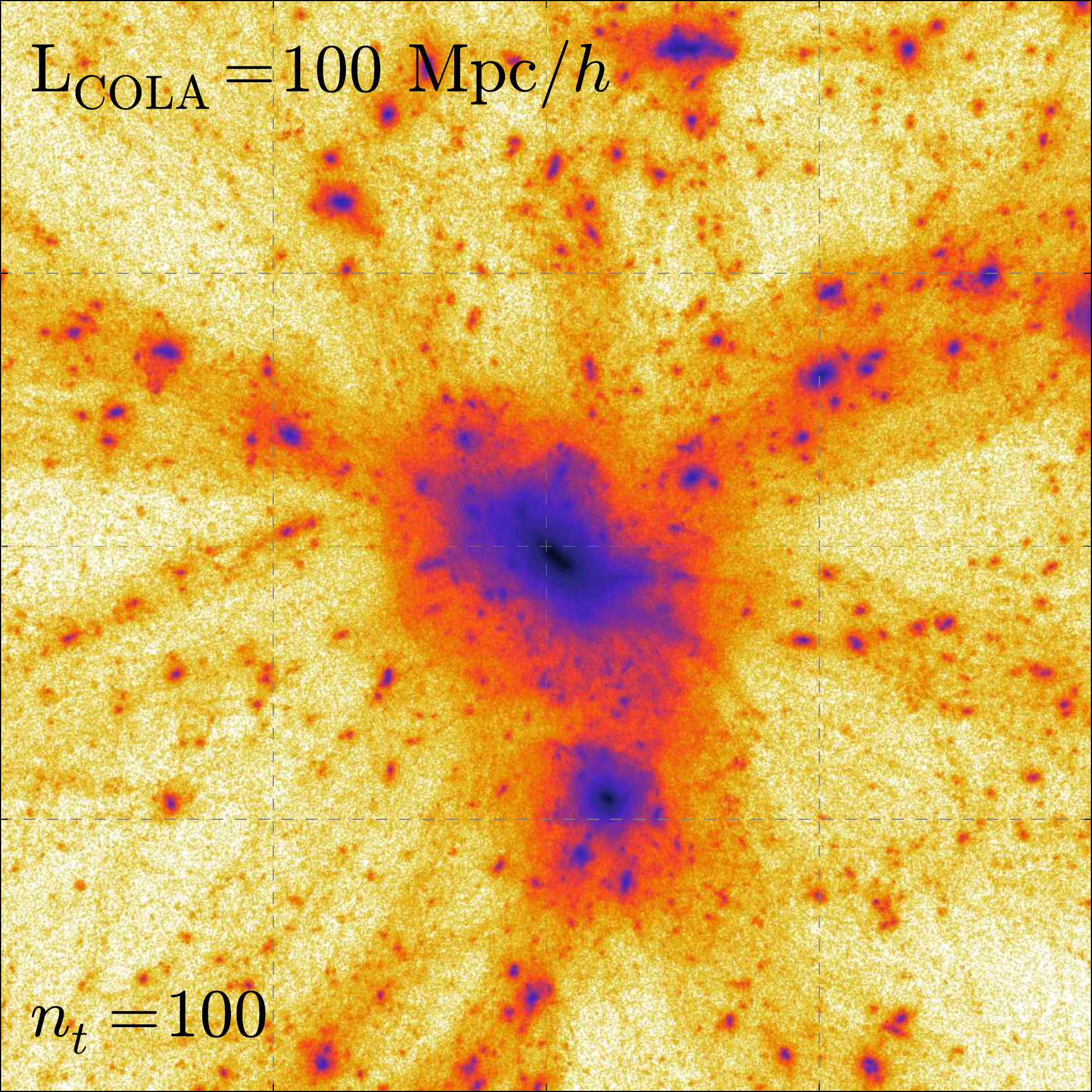}}
\subfloat{\includegraphics[width=0.248\textwidth]{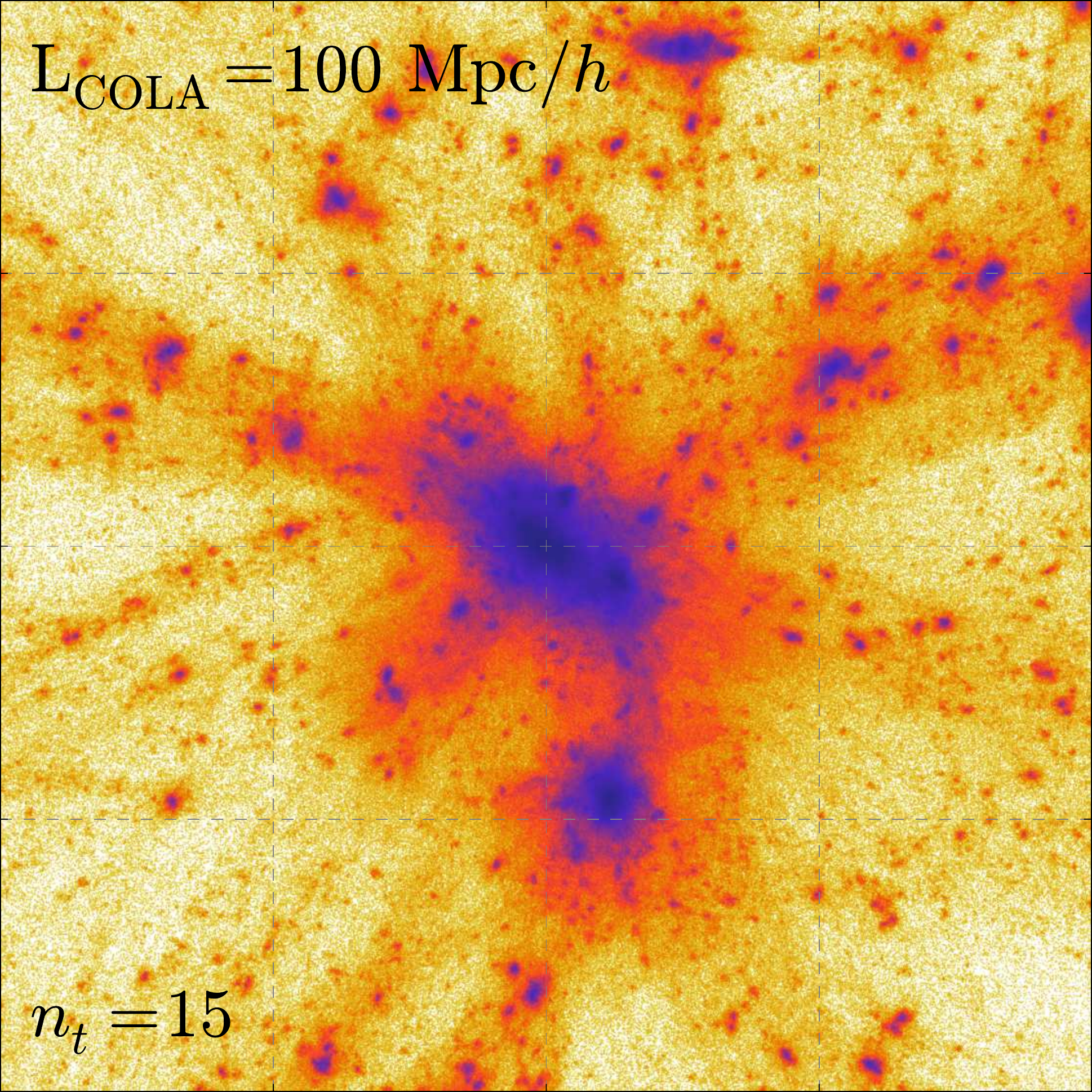}}
\subfloat{\includegraphics[width=0.248\textwidth]{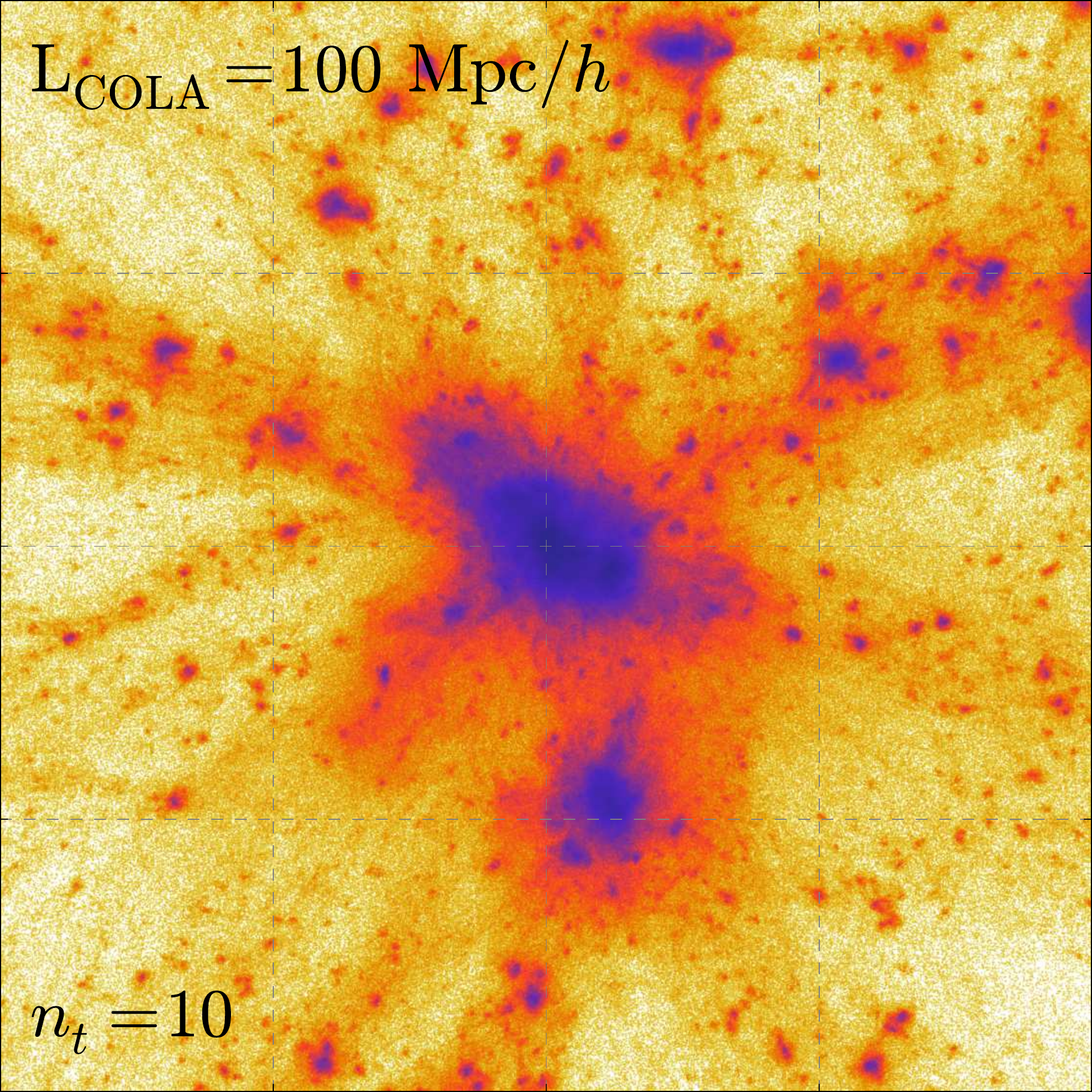}}
\subfloat{\includegraphics[width=0.248\textwidth]{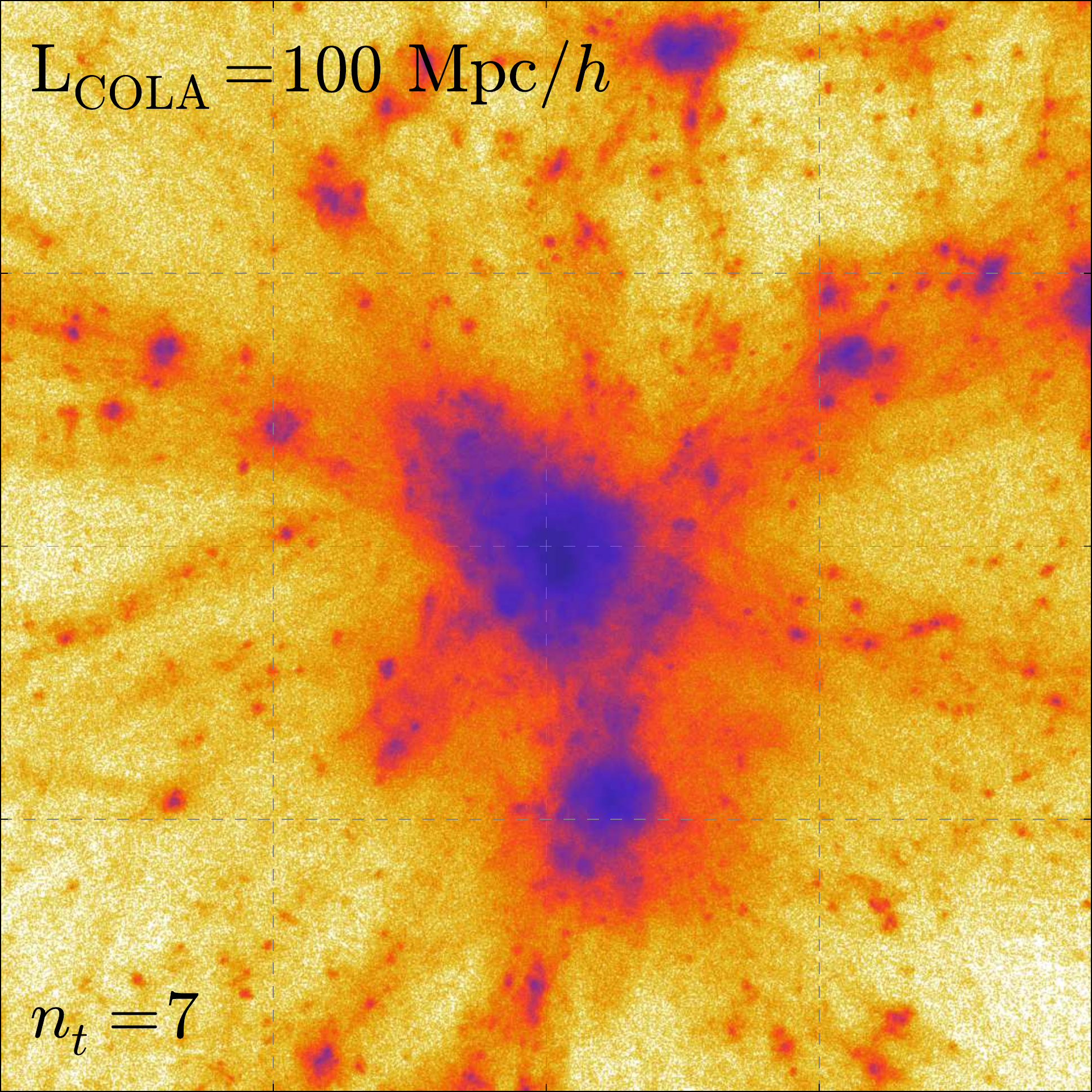}}
\hfill
\caption{Slices of thickness 12.5Mpc$/h$ at $z=0$ through one and the same CDM density field as obtained with the original time-domain-only COLA method by varying the number of timesteps, $n_t$, across columns. Rows show the result at different magnifications, with the top row slices being 25Mpc$/h$ on the side, and the bottom row slices  being 12.5Mpc$/h$ on the side. One should note that for $n_t\lesssim10$ halos puff up significantly, rendering their identification more unreliable. We use the same color map for all projected
density snapshots in this paper.} \label{fig:slices}
\end{figure}

The second order initial  conditions (giving $\x_{\mathrm{LPT}}$) in the full box are calculated\footnote{The adopted cosmology in standard notation is: $\Omega_M= 0.274$,
$\Omega_\Lambda= 0.726$, $\Omega_b= 0.046$,  $h=0.7$, $\sigma_8=0.8$, $n_s= 0.95$.
} with  MUSIC \hfill using \hfill $512^3$ \hfill particles \hfill in \hfill the \hfill full \hfill box of \hfill size \hfill $L=100$\,Mpc$/h$. \hfill The \hfill central \hfill 45\% \begin{figure}[H]
\centering
\subfloat{\includegraphics[width=0.248\textwidth]{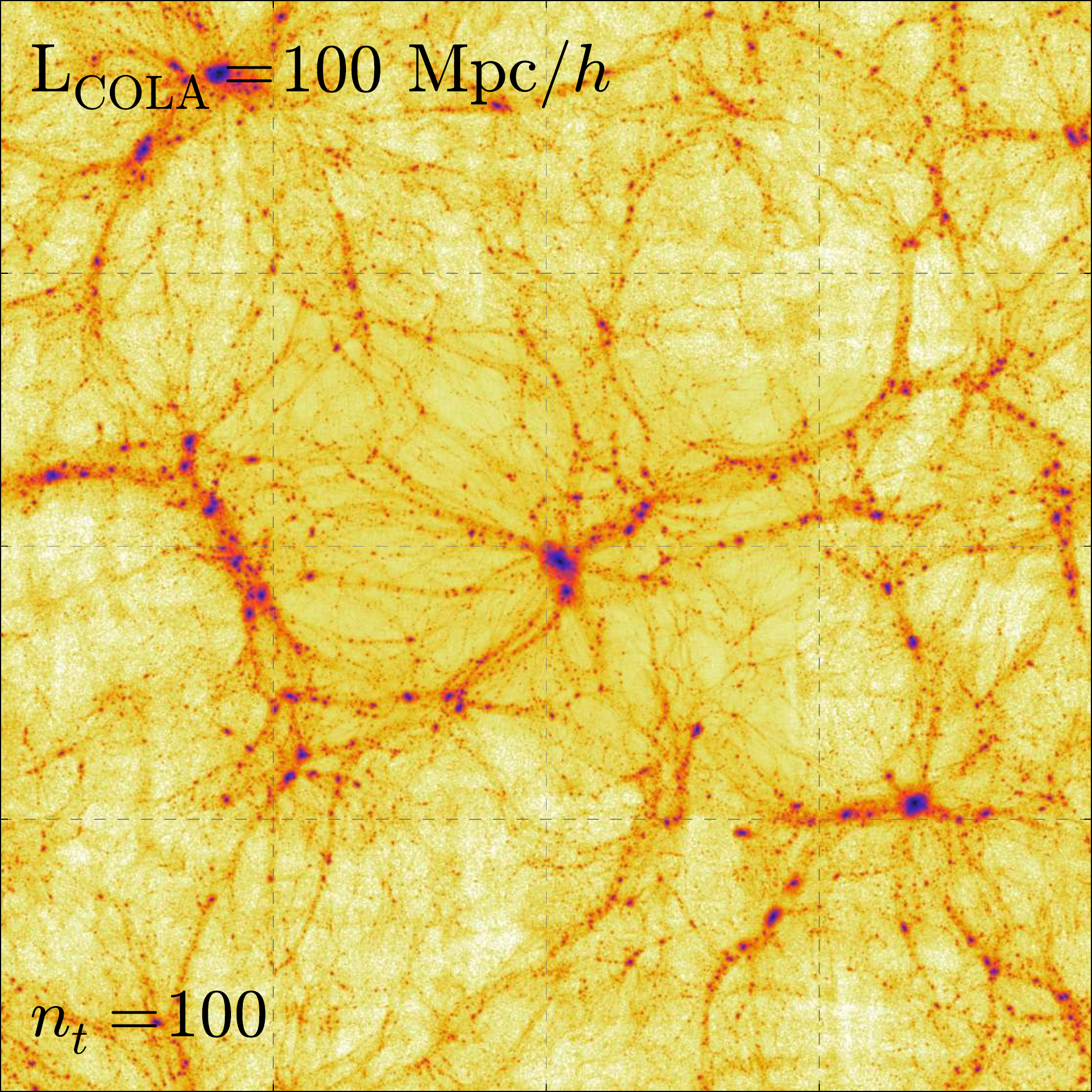}}
\subfloat{\includegraphics[width=0.248\textwidth]{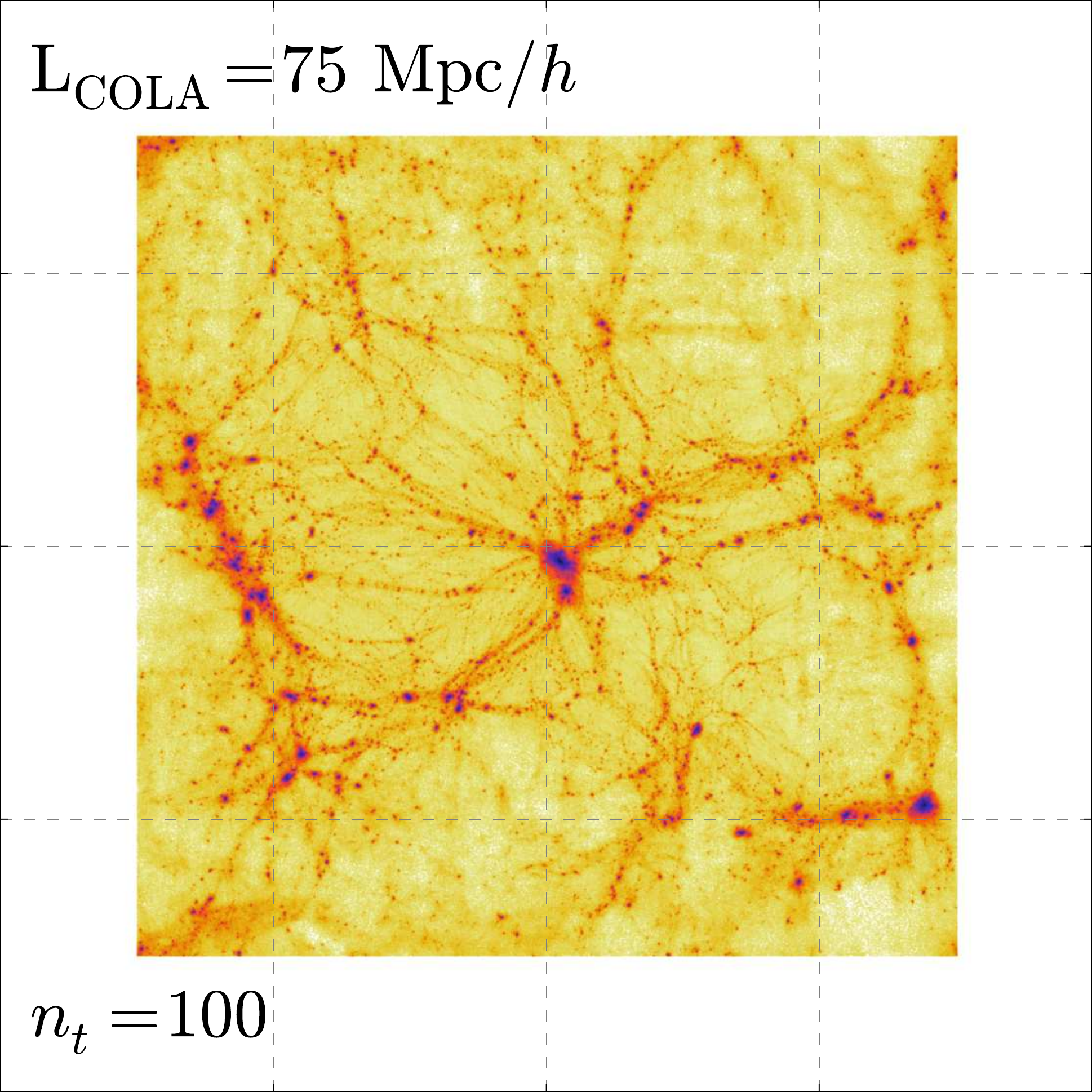}}
\subfloat{\includegraphics[width=0.248\textwidth]{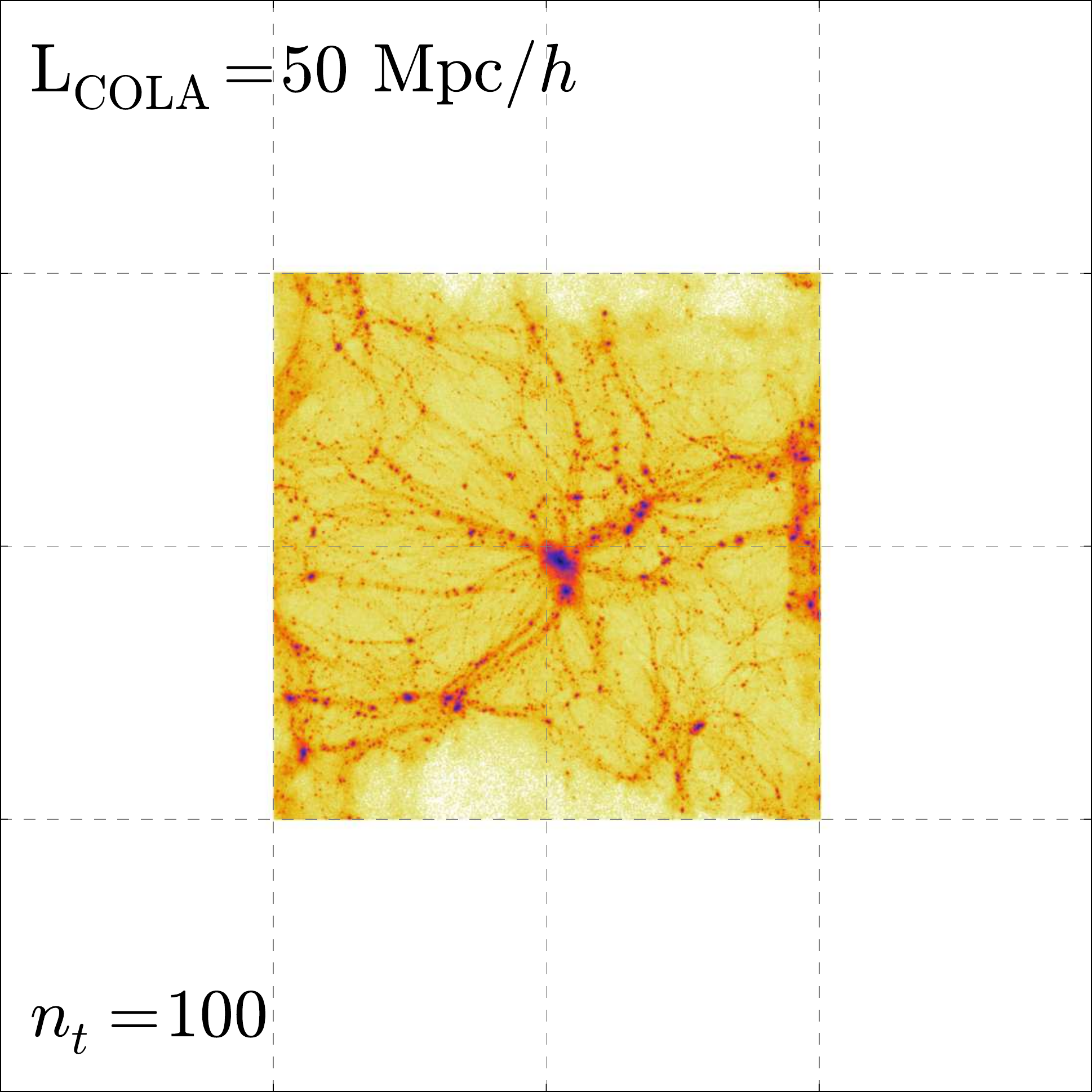}}
\subfloat{\includegraphics[width=0.248\textwidth]{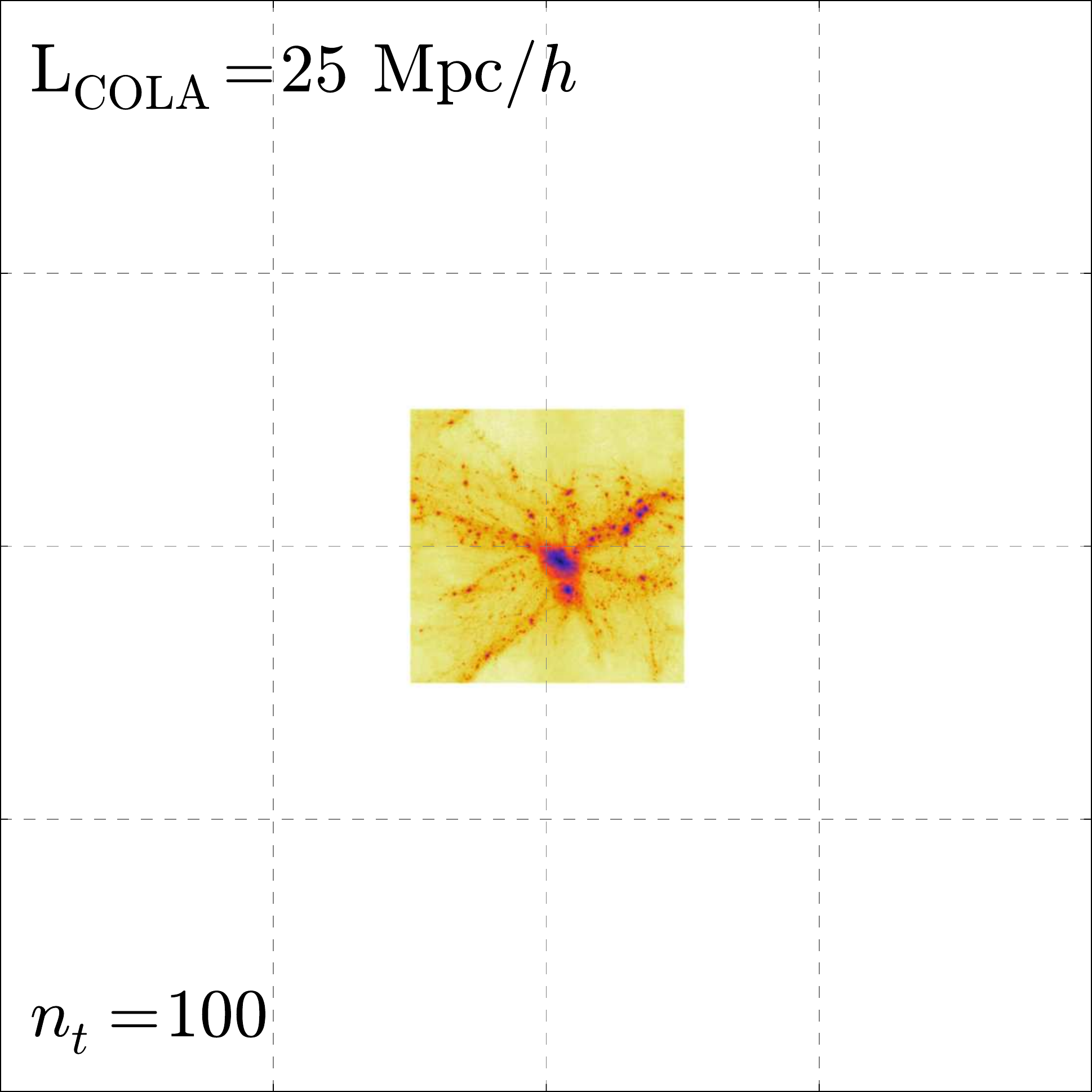}}
\hfill
\\
\vspace{-0.38cm}
\subfloat{\includegraphics[width=0.248\textwidth]{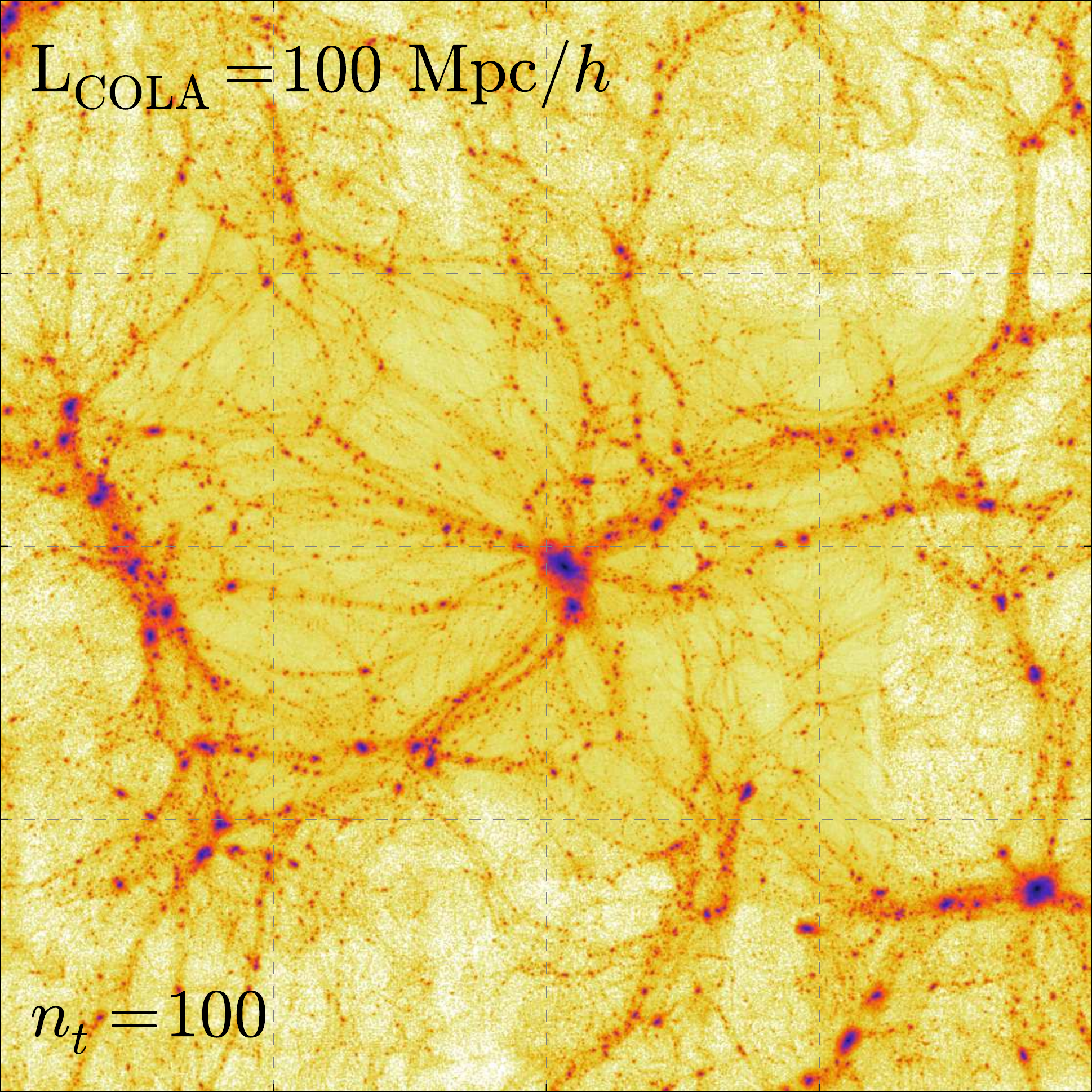}}
\subfloat{\includegraphics[width=0.248\textwidth]{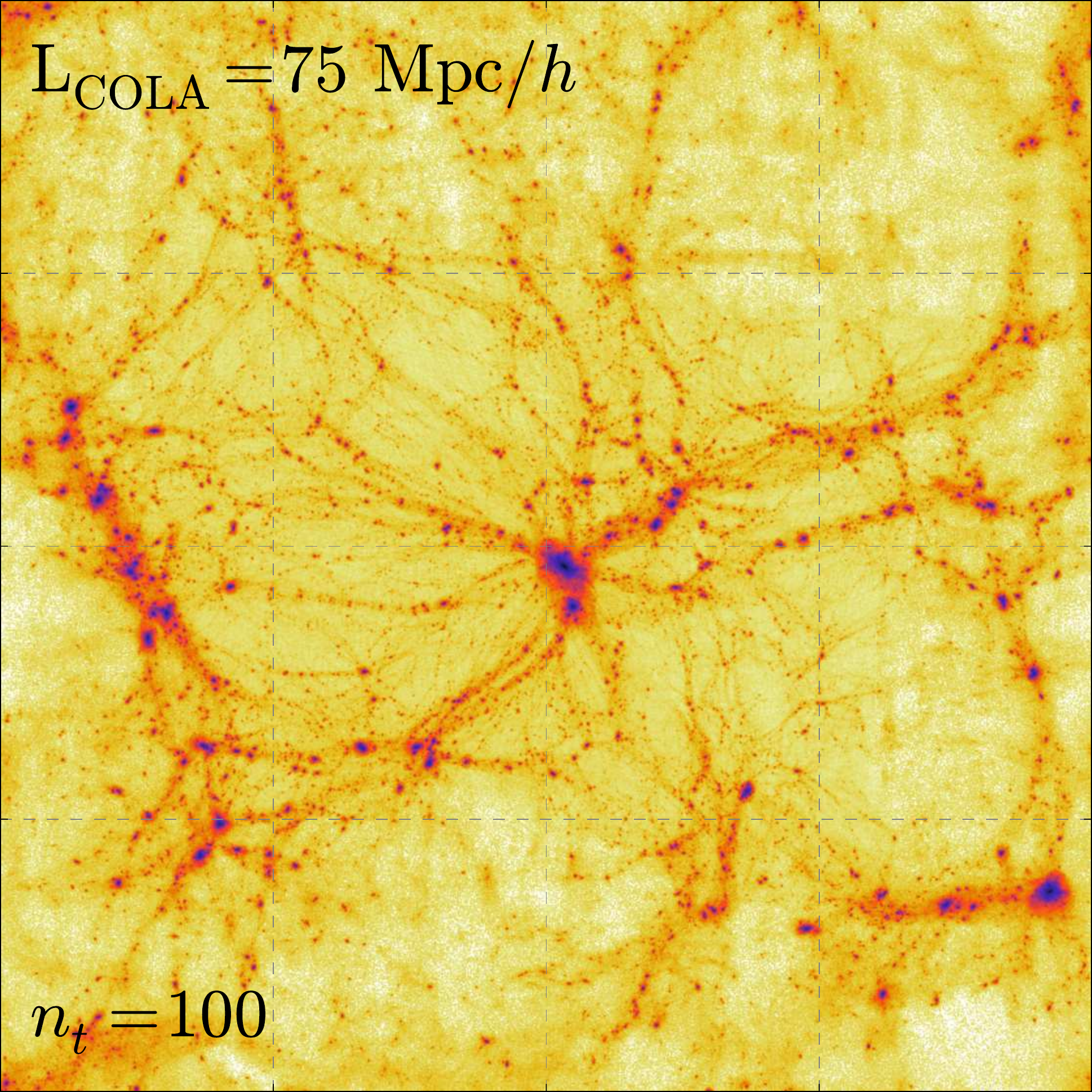}}
\subfloat{\includegraphics[width=0.248\textwidth]{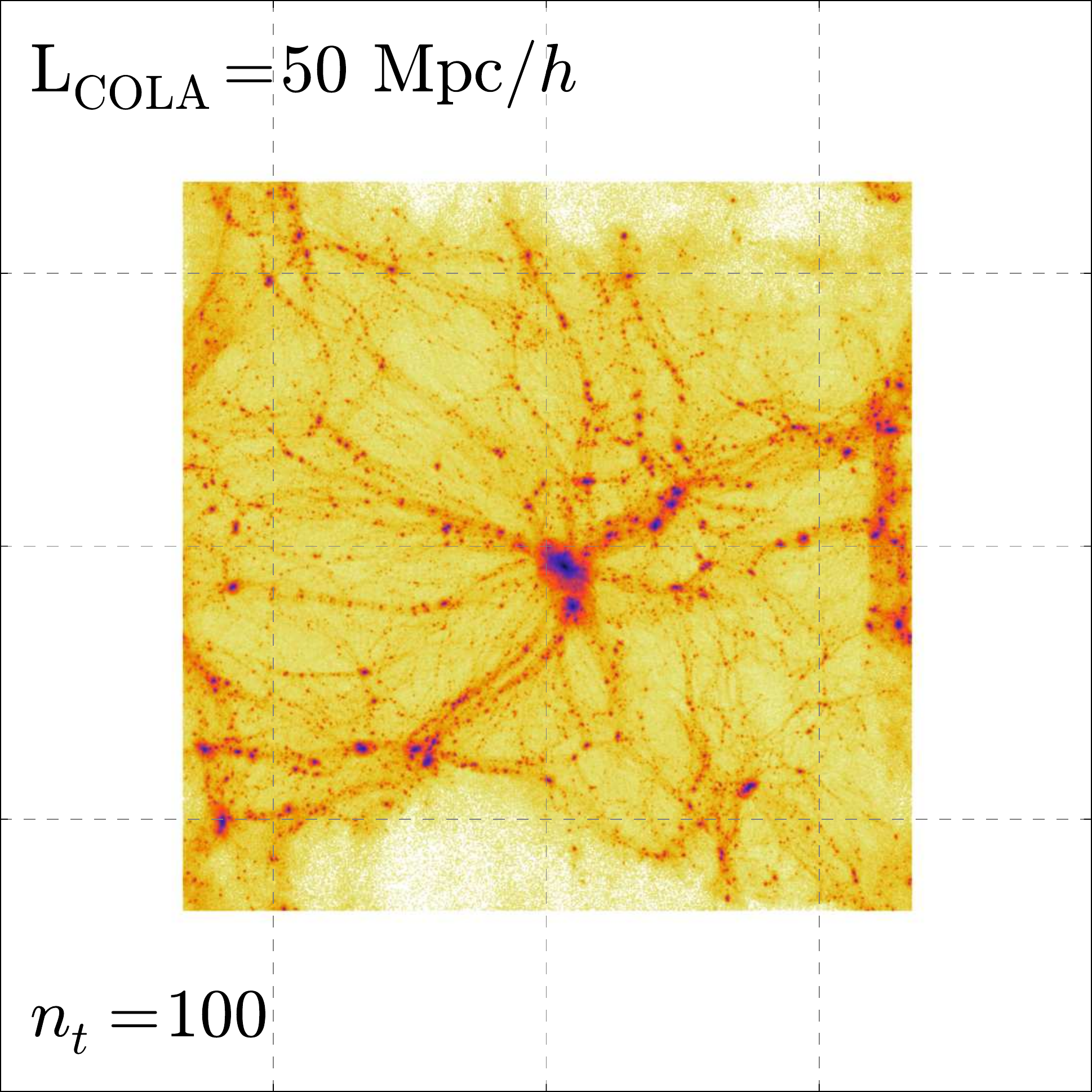}}
\subfloat{\includegraphics[width=0.248\textwidth]{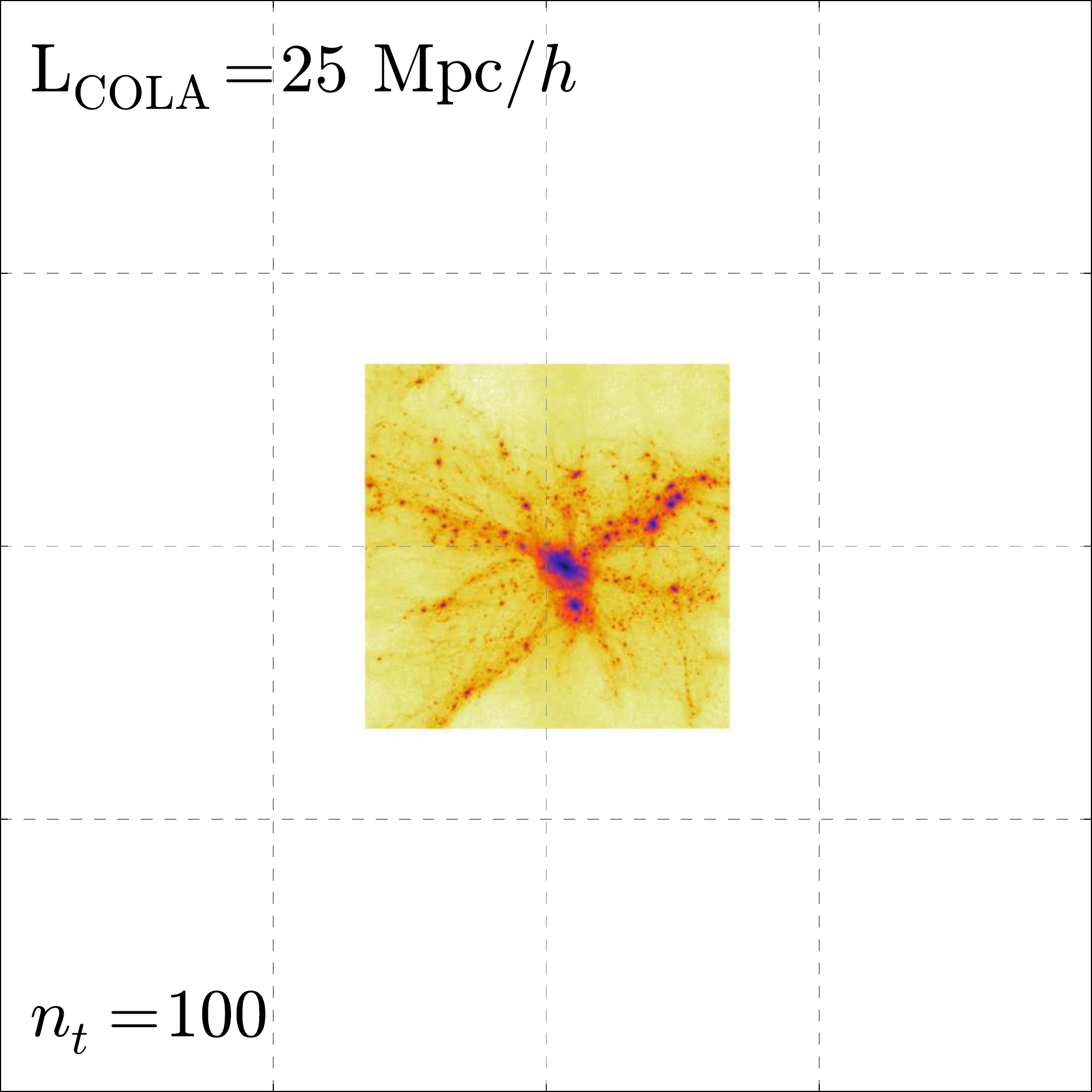}}
\hfill
\\
\vspace{-0.38cm}
\subfloat{\includegraphics[width=0.248\textwidth]{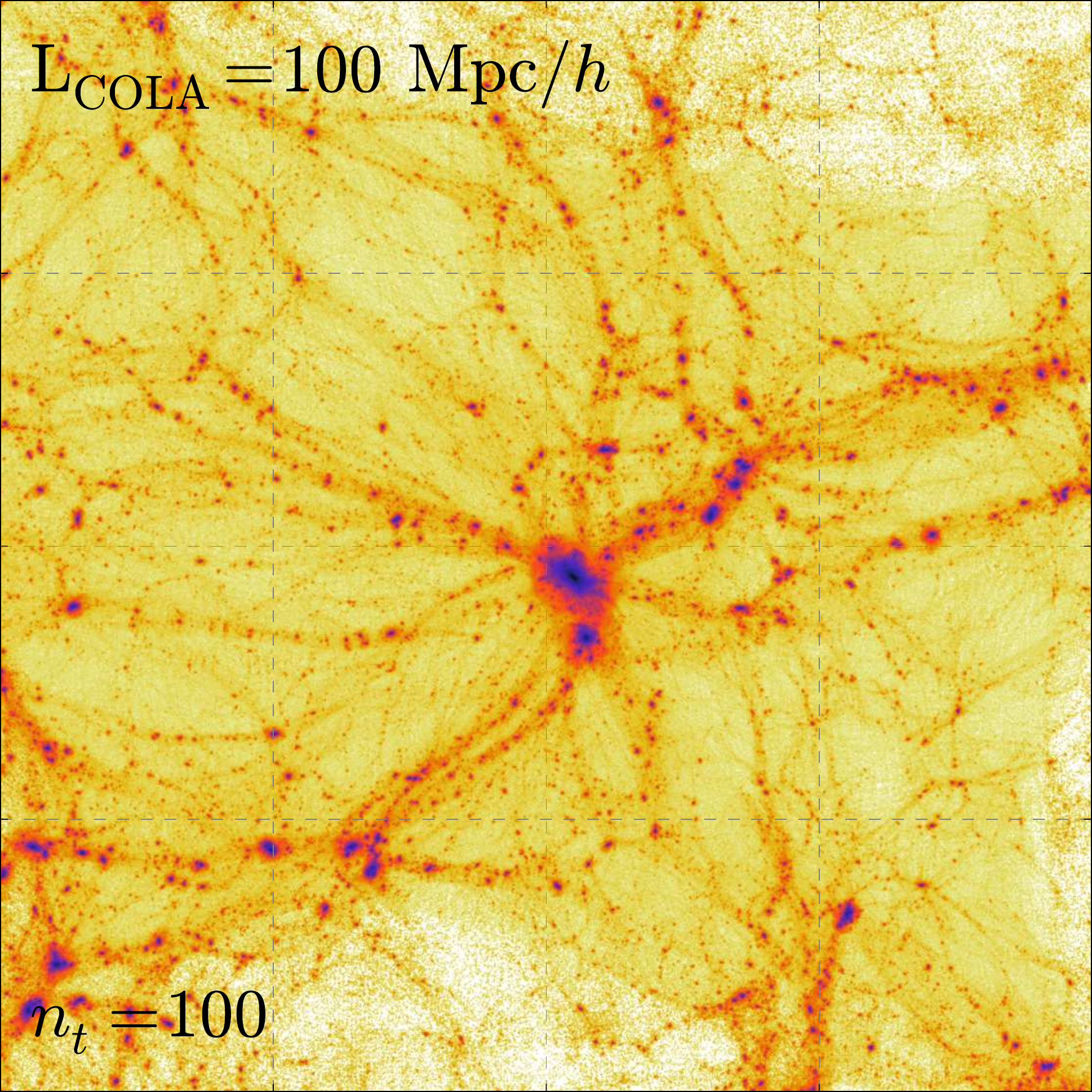}}
\subfloat{\includegraphics[width=0.248\textwidth]{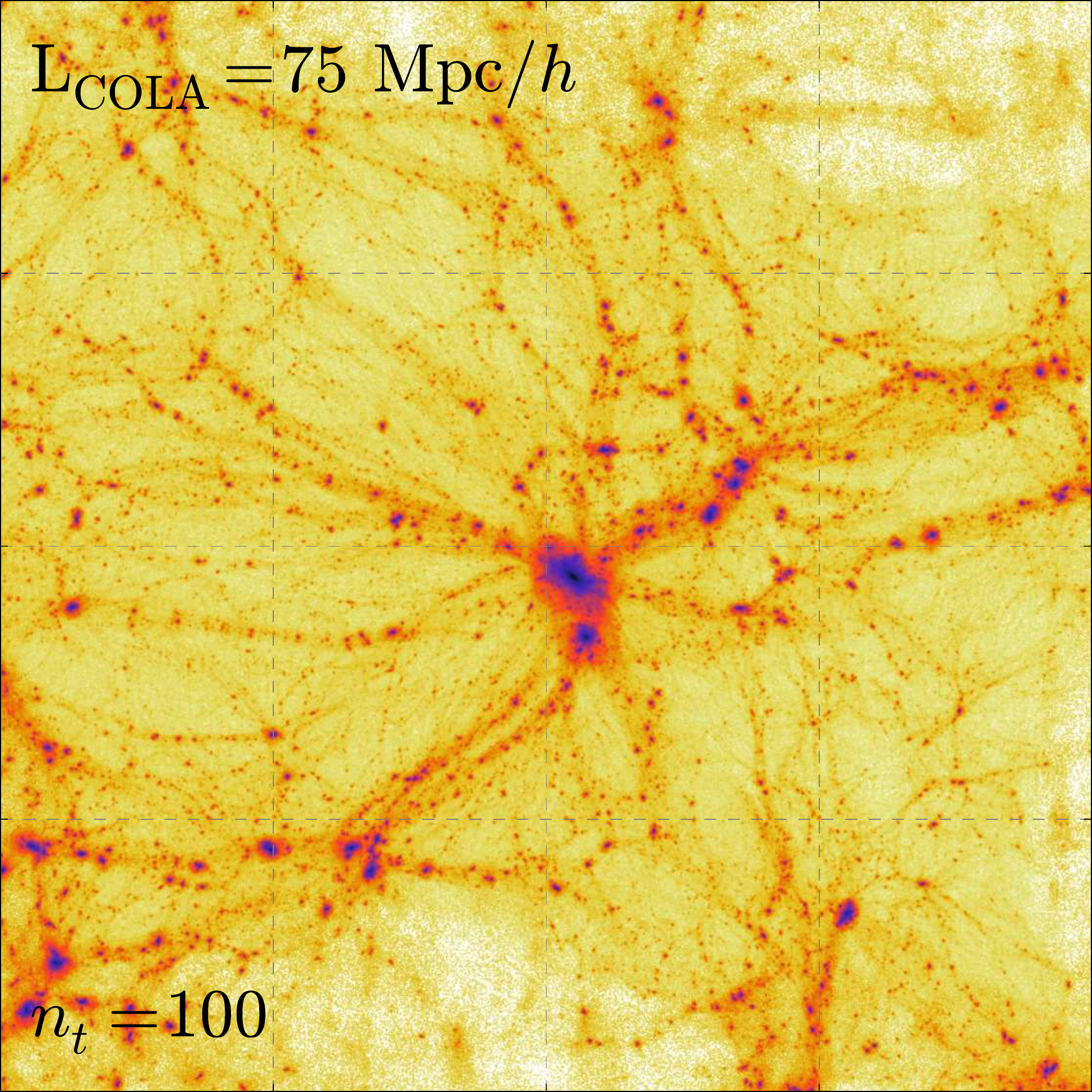}}
\subfloat{\includegraphics[width=0.248\textwidth]{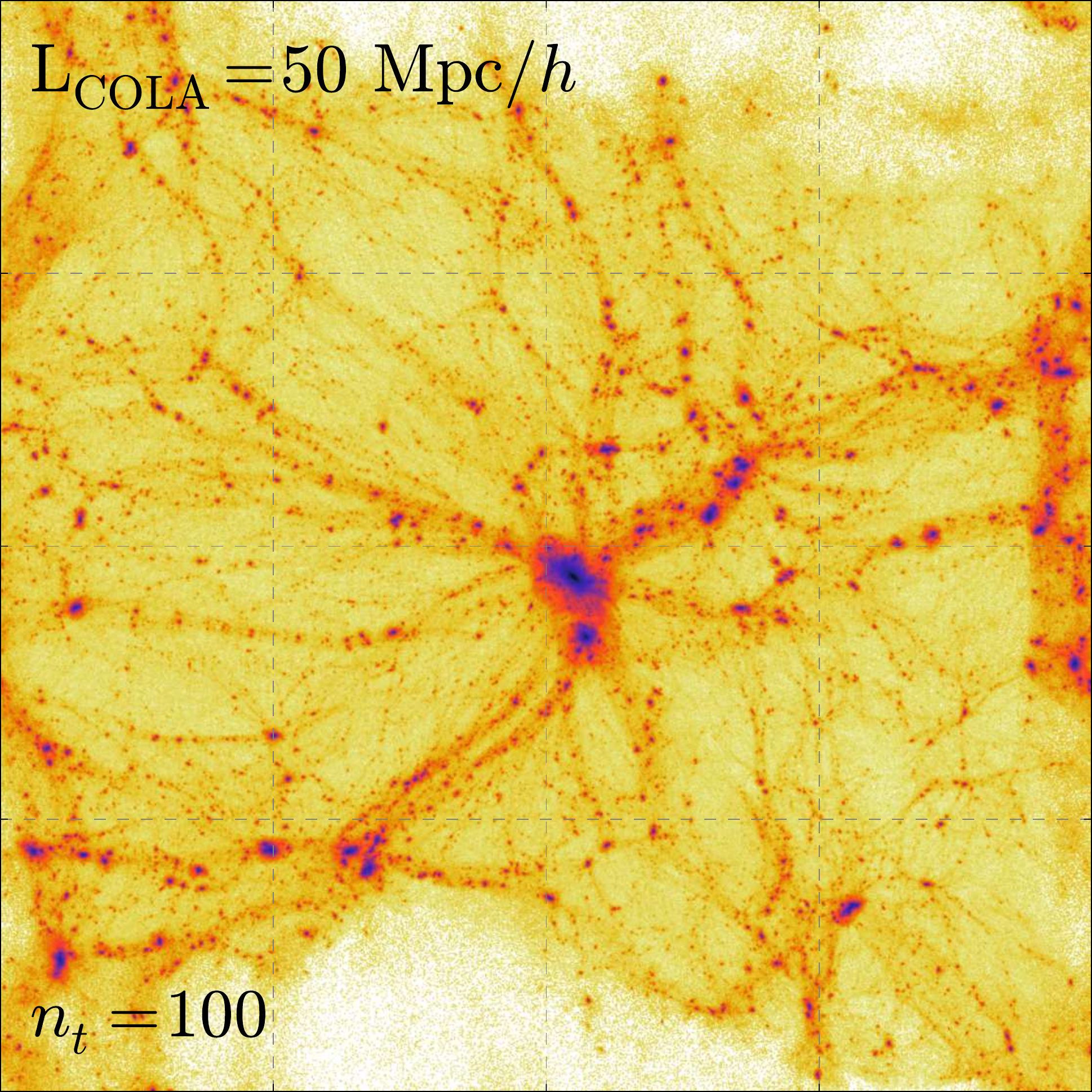}}
\subfloat{\includegraphics[width=0.248\textwidth]{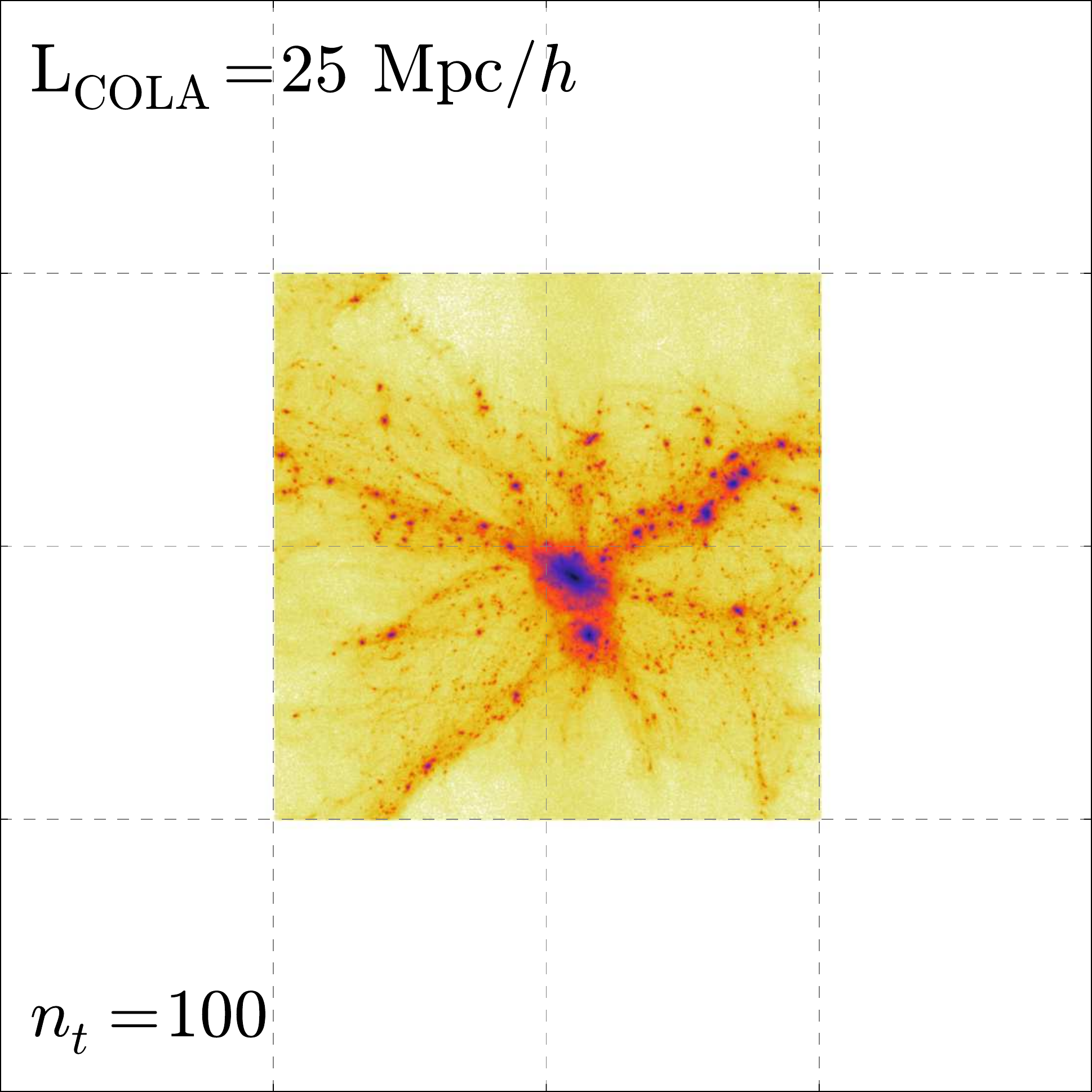}}
\hfill
\\
\vspace{-0.38cm}
\subfloat{\includegraphics[width=0.248\textwidth]{LAST_full_n100_cut0_L25.pdf}}
\subfloat{\includegraphics[width=0.248\textwidth]{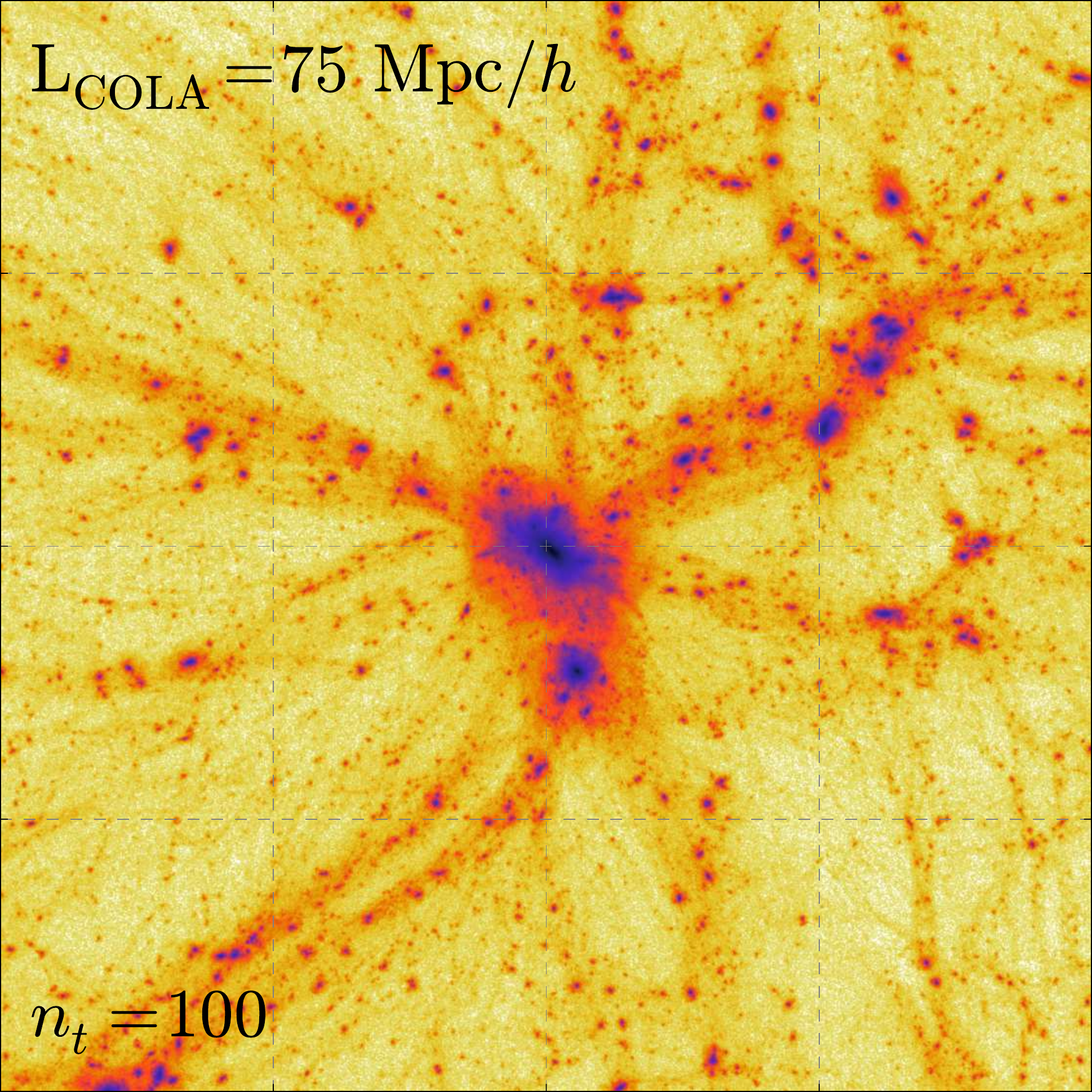}}
\subfloat{\includegraphics[width=0.248\textwidth]{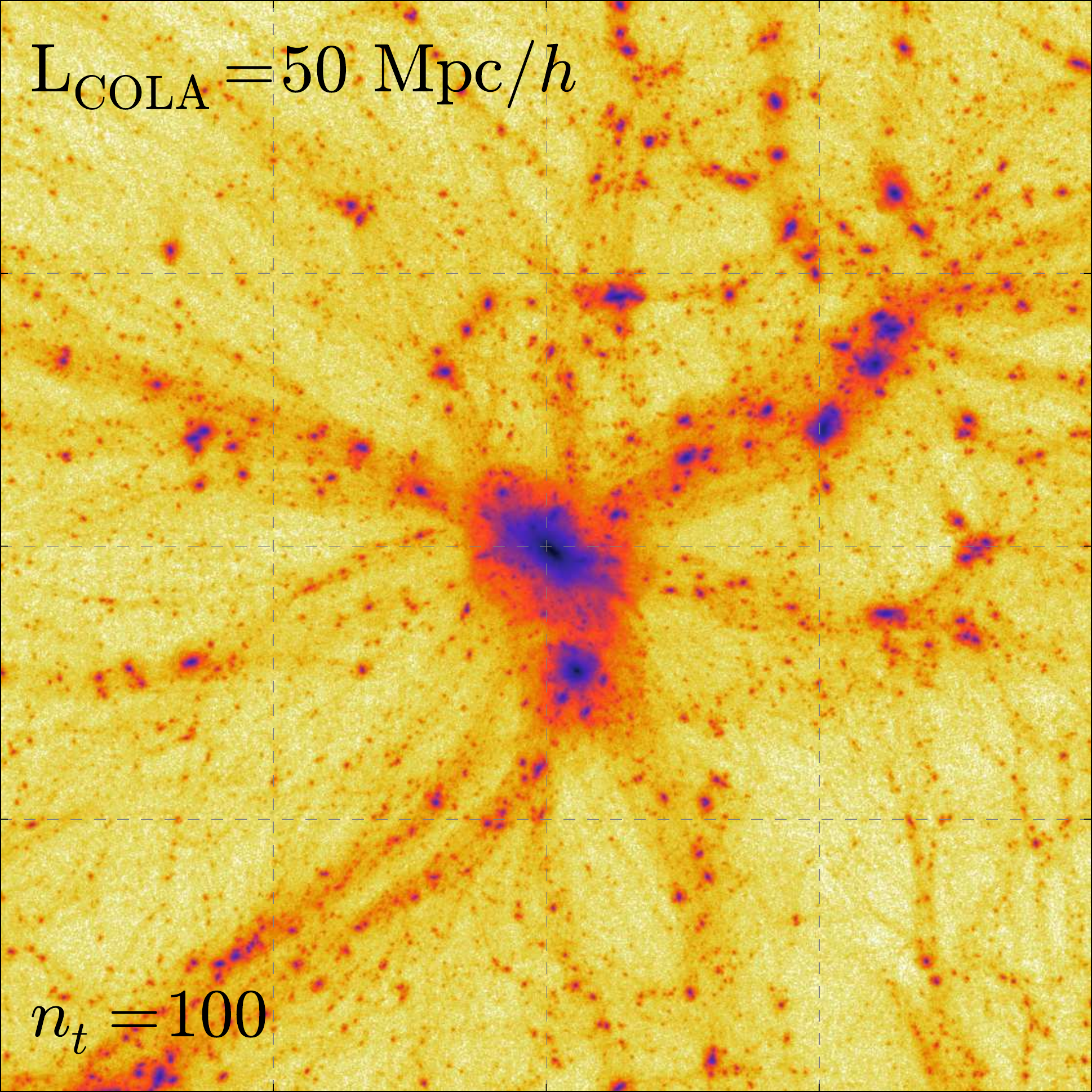}}
\subfloat{\includegraphics[width=0.248\textwidth]{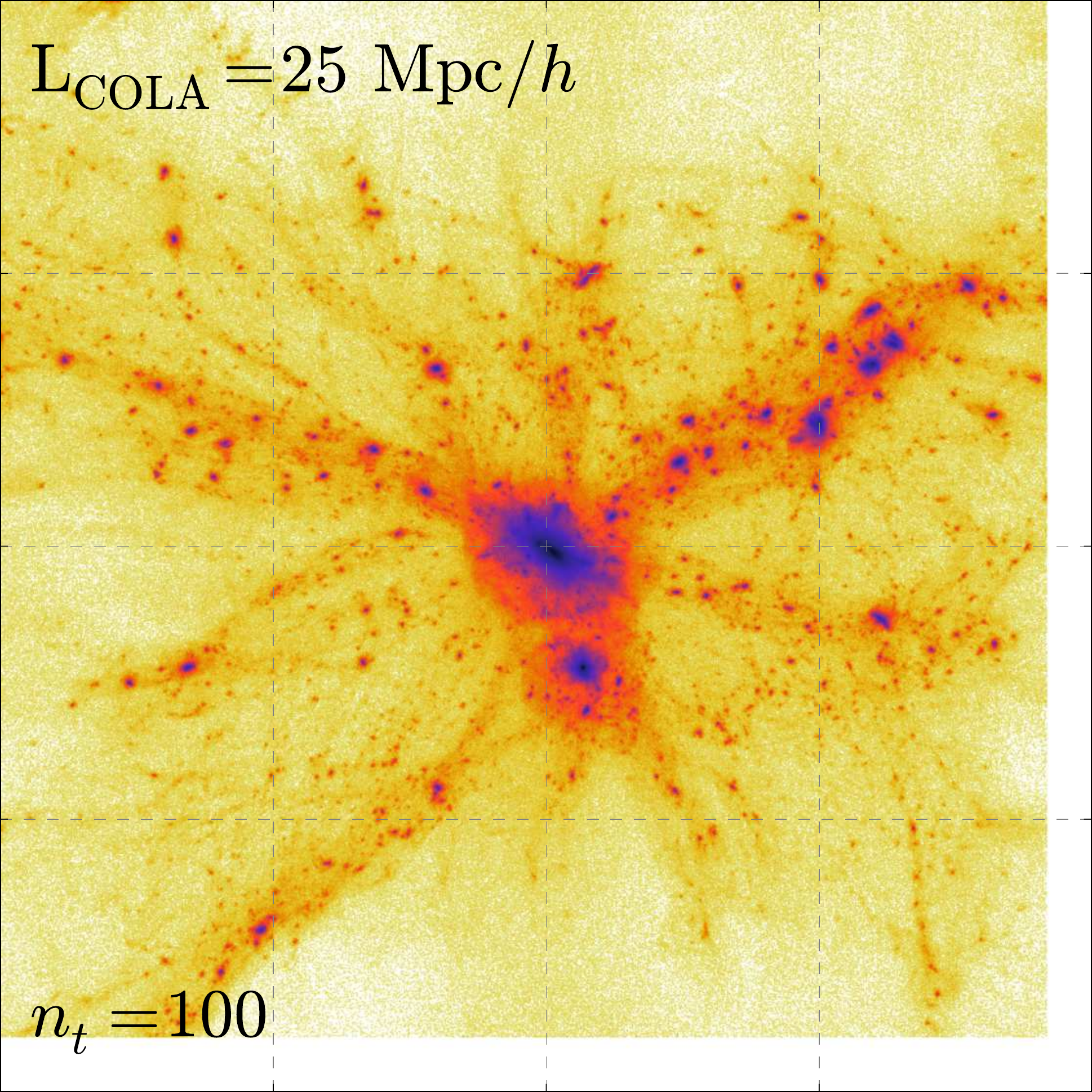}}
\hfill
\\
\vspace{-0.38cm}
\subfloat{\includegraphics[width=0.248\textwidth]{LAST_full_n100_cut0_L12_5.pdf}}
\subfloat{\includegraphics[width=0.248\textwidth]{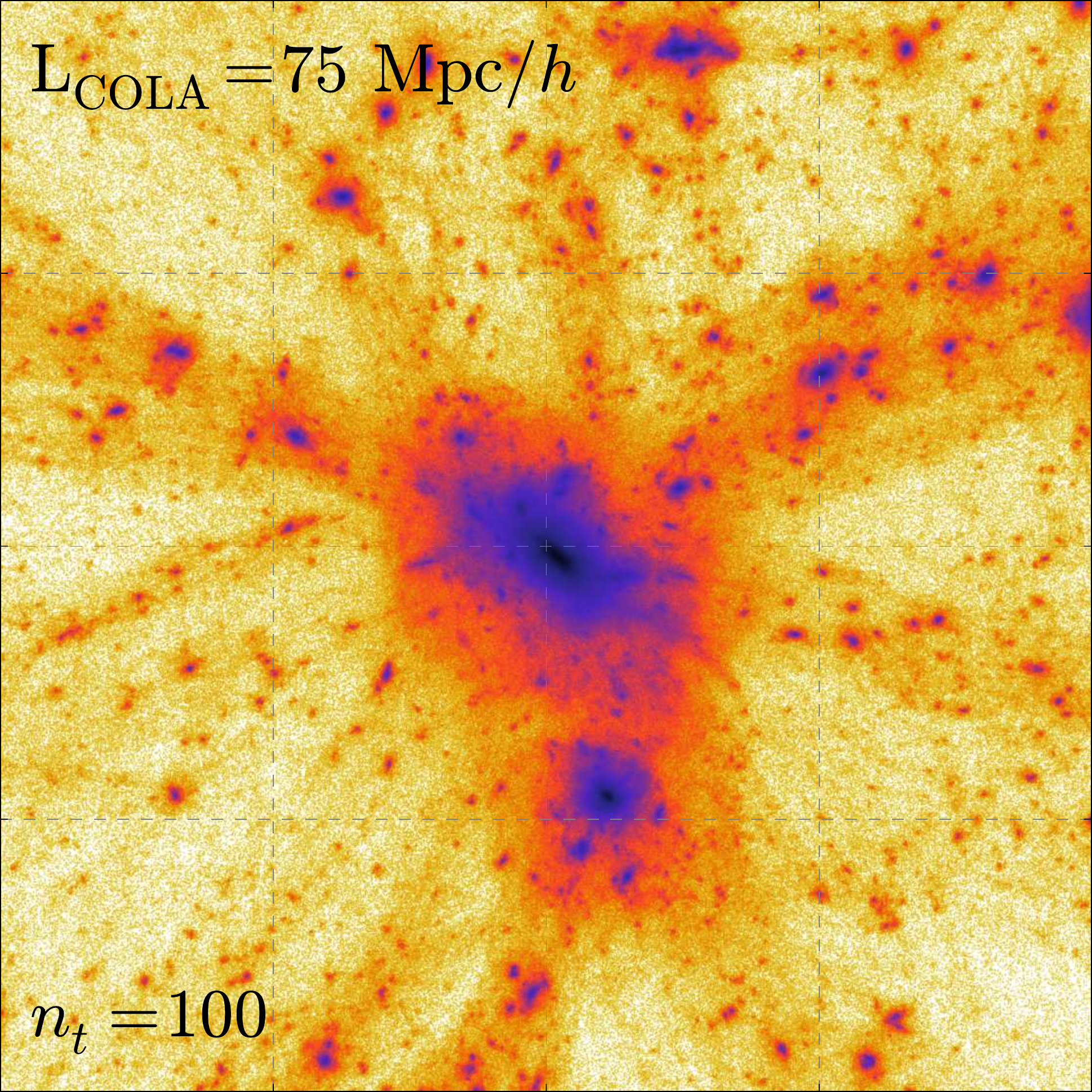}}
\subfloat{\includegraphics[width=0.248\textwidth]{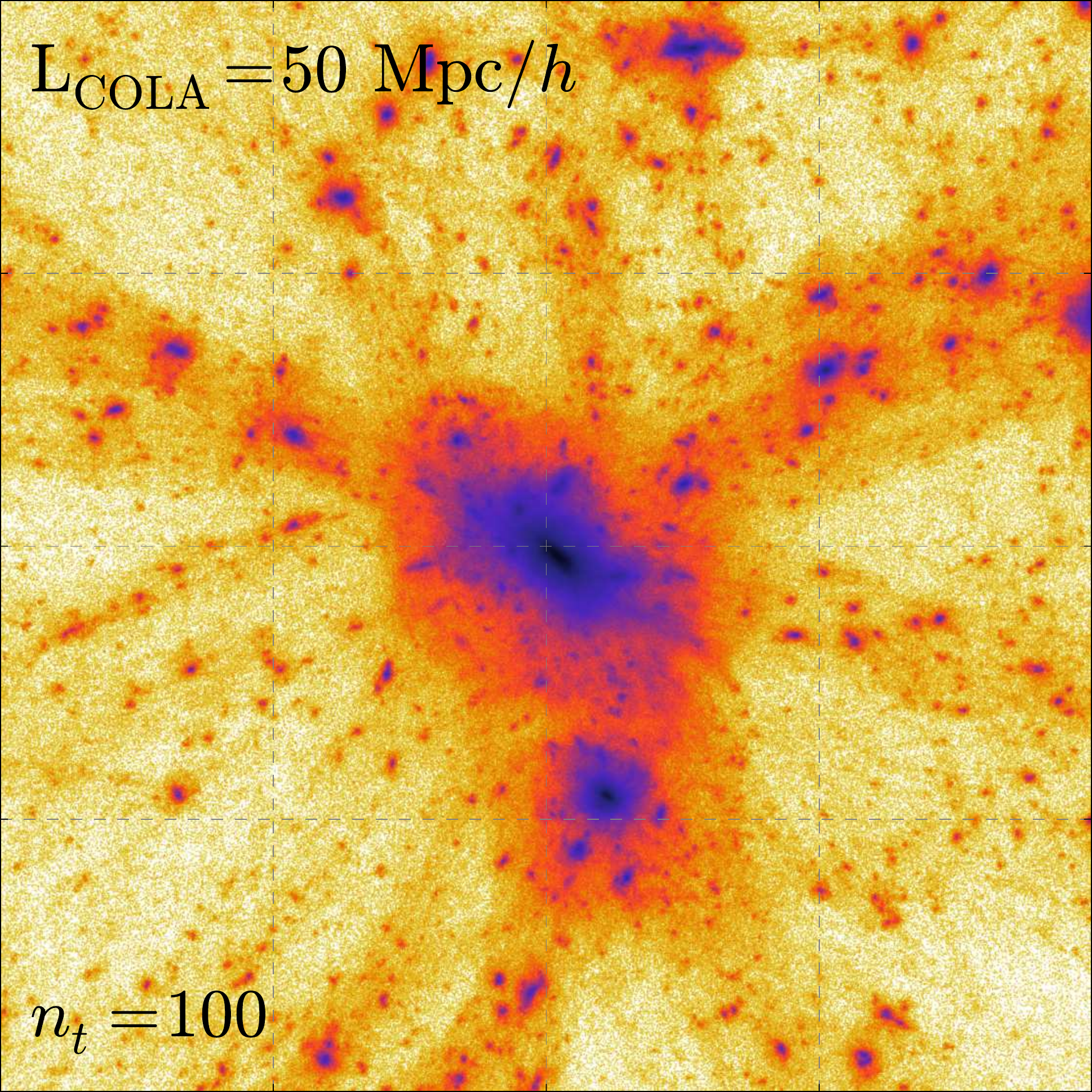}}
\subfloat{\includegraphics[width=0.248\textwidth]{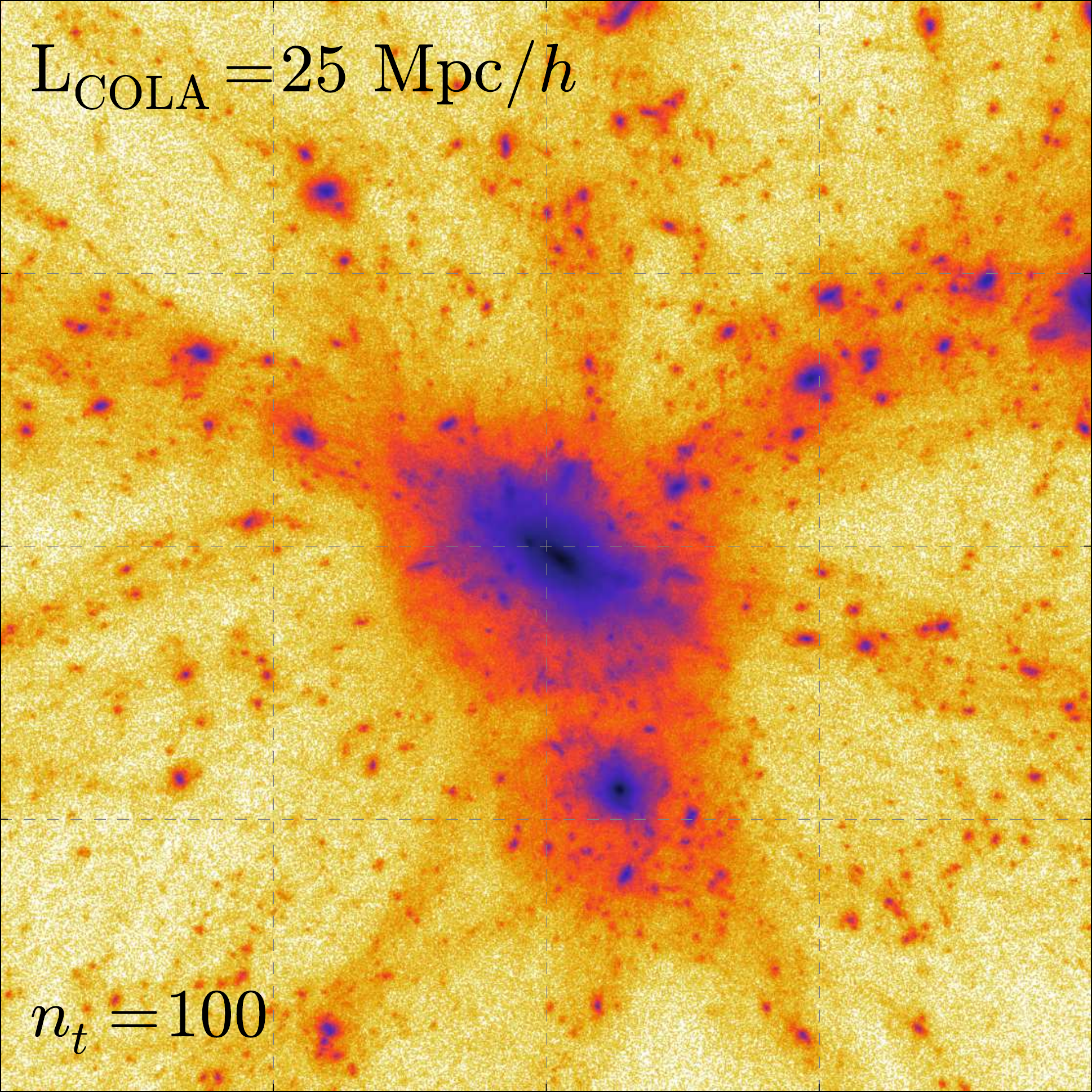}}
\hfill
\caption{Slices of thickness 12.5Mpc$/h$ at $z=0$ through the same CDM density field as obtained with the proposed extension of the COLA method to the spatial domain. The number of timesteps for all slices is fixed at $n_t=100$. Columns show different choices of $\mathrm{L}_{\mathrm{COLA}}$.  Rows show the result at consecutive magnifications, with the slices being 100 (top row), 75, 50, 25, 12.5 (bottom row) \,Mpc$/h$ on the side. For $\mathrm{L}_{\mathrm{COLA}} = 25$\,Mpc$/h$, not all particles in the vicinity of the central halo are present in the initial Lagrangian volume, which leads to noticeable differences in the distribution of substructures. Note that the slices in the last two rows are offset  relative to the other rows, so that they are centered around the most massive halo.} \label{fig:slices100}
\end{figure}\noindent(lengthwise) of the full box are refined with a grid of particles with a particle spacing of $L/1024=0.098$\,Mpc$/h$, corresponding to 
particle masses  of $7.08\times 10^7$\,M$_\odot/h$. We evolved those same initial conditions with pyCOLA  from redshift of 9 down to redshift of zero using various number of timesteps, $n_t$, and COLA
 box sizes, L$_\mathrm{COLA}$. The LPT displacement field inside the COLA volume was calculated to second order with the 2-pt rules in Appendix~\ref{app:KDKlpt} using pyCOLA. We then evolved the particle positions using pyCOLA with $n_t=100,15,10,7$ and L$_\mathrm{COLA}=100,75,50,25$\,Mpc$/h$. The simulations were centered around a halo of mass $2.08\times 10^{14}$\,M$_\odot/h$. We used a fixed force resolution (PM grid spacing) of $L/1536=0.065$\,Mpc$/h$ for all runs. Note that both the size of the PM grid and the number of particles inside the COLA volume change with L$_\mathrm{COLA}$ as
 we are keeping the force resolution and particle masses constant, while changing the volume (corresponding to L$_\mathrm{COLA}$) simulated by the N-body code.

In Figure~\ref{fig:slices} we show slices through the CDM density field obtained with the original temporal COLA method\footnote{Note that we did not try to optimize the choice for $n_{\mathrm{LPT}}$ for this simulation set-up as prescribed in TZE, so the results we present here  using the original COLA method can be suboptimal.} by setting (see Section~\ref{scola:disc}) the full box and COLA box sizes to be the same, i.e. $L=$L$_\mathrm{COLA}=100$\,Mpc$/h$. We vary the number of timesteps between 100 and 7 across the columns of that panel plot. The two rows correspond to two different magnifications: the top slices are $25$\,Mpc$/h$ on the side, while the bottom are $12.5$\,Mpc$/h$. One can clearly see that the original temporal COLA method behaves as expected. It reproduces the large-scale features  faithfully, while performing gradually worse at small scales when $n_t$ is lowered. For $n_t\lesssim10$ halos puff up significantly as the few timesteps become insufficient for the N-body side of the code to keep particles tightly bound  to halos (see TZE). One can see that from Table~\ref{table:nt} (discussed in more detail below) as well, where the central halo mass shows a gradual decline with decreasing $n_t$, while the halo size increases. Note that the errors in the halo position in both real and redshift space are within $500$\,kpc$/h$, while the finger-of-god (FoG) effect is consistently underestimated. For a discussion of how these imperfections can be fixed, see TZE for example.

\begin{figure}[p]
\centering
\subfloat{\includegraphics[width=0.248\textwidth]{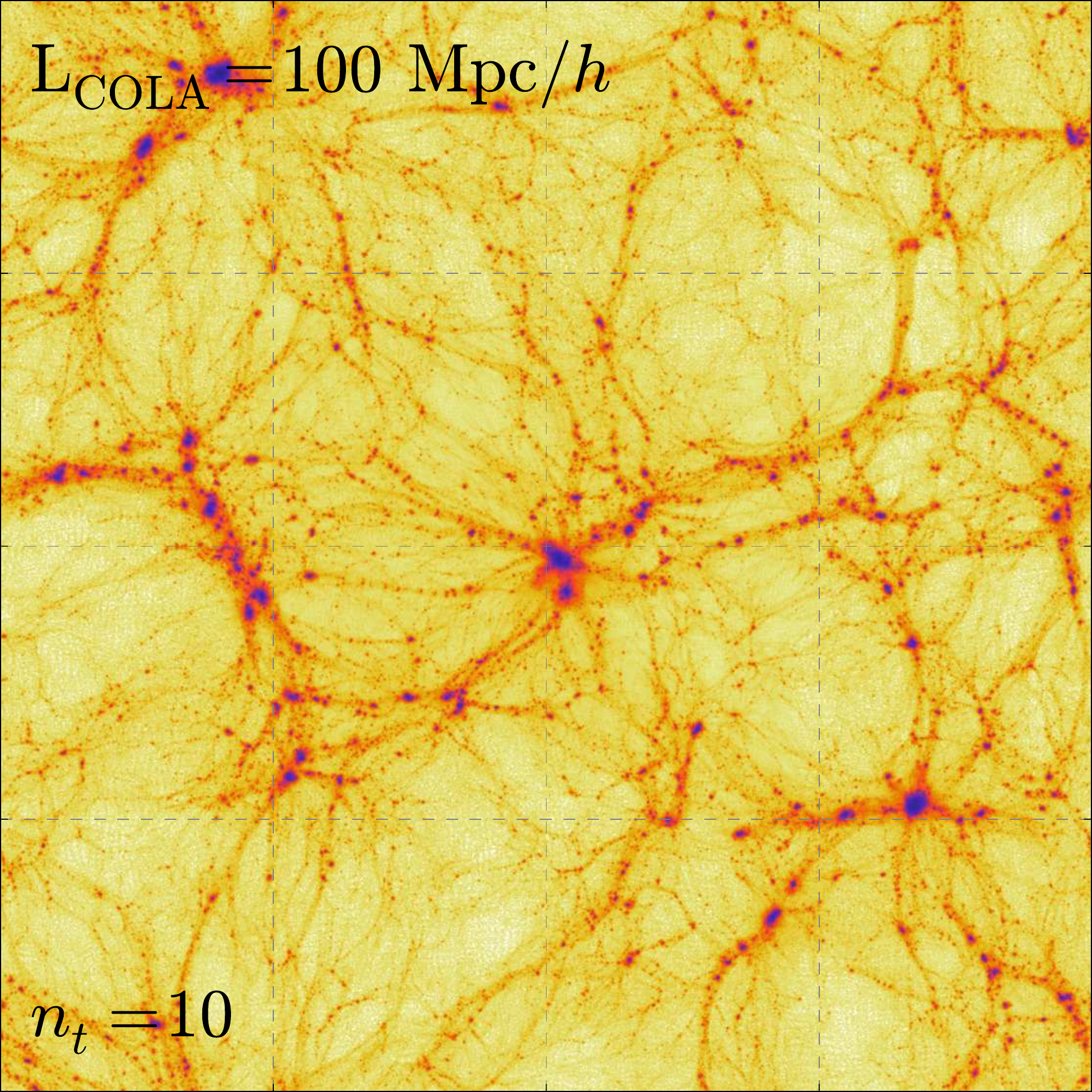}}
\subfloat{\includegraphics[width=0.248\textwidth]{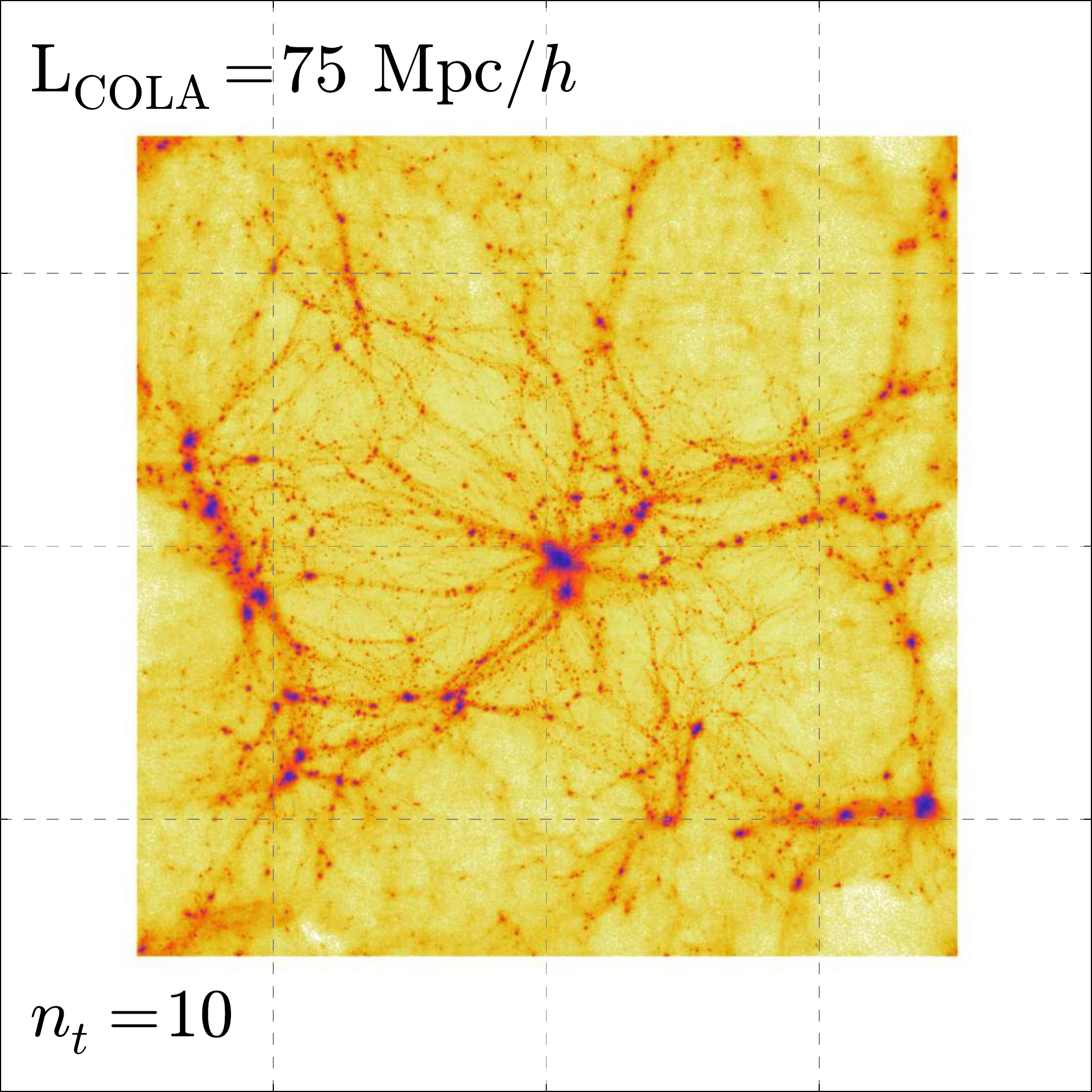}}
\subfloat{\includegraphics[width=0.248\textwidth]{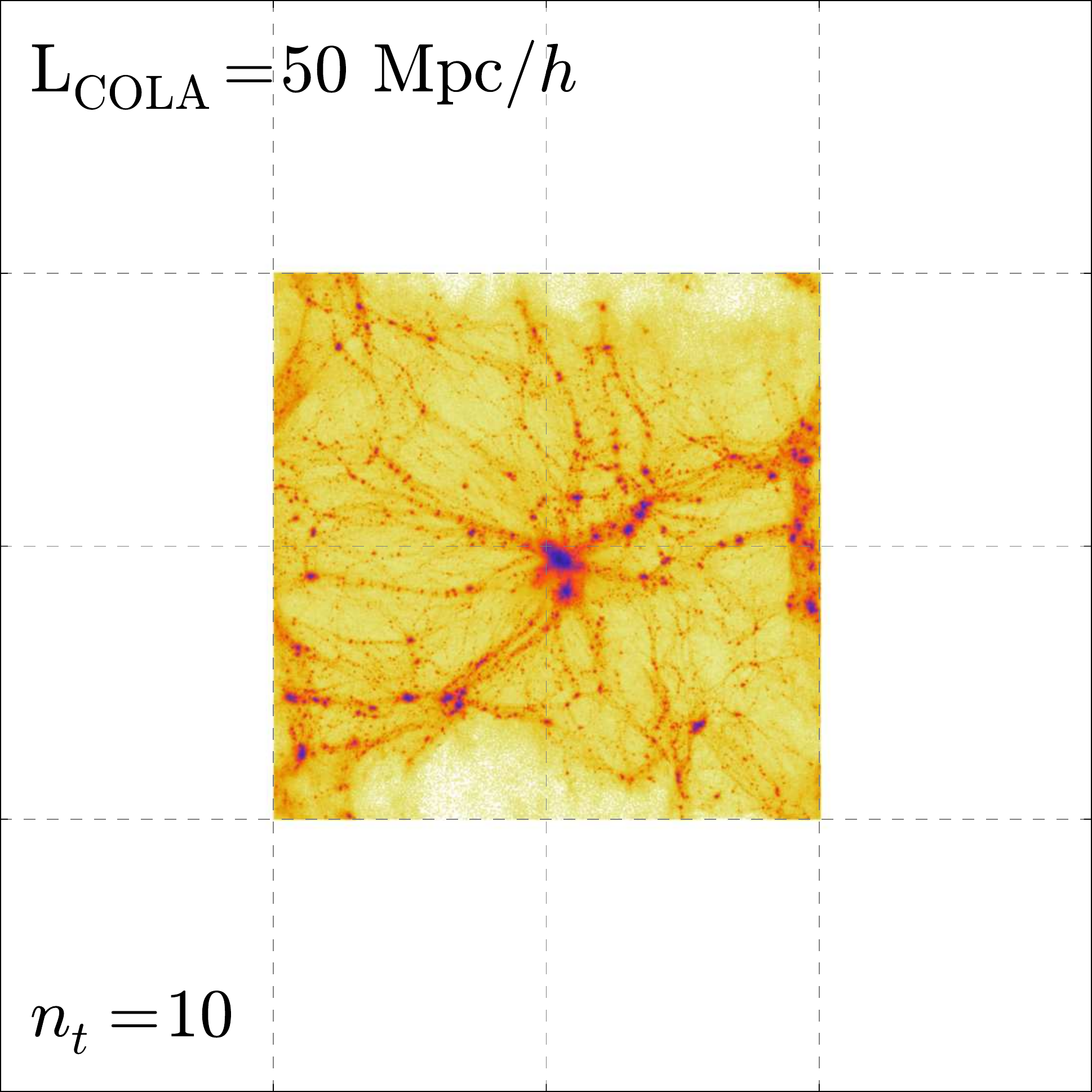}}
\subfloat{\includegraphics[width=0.248\textwidth]{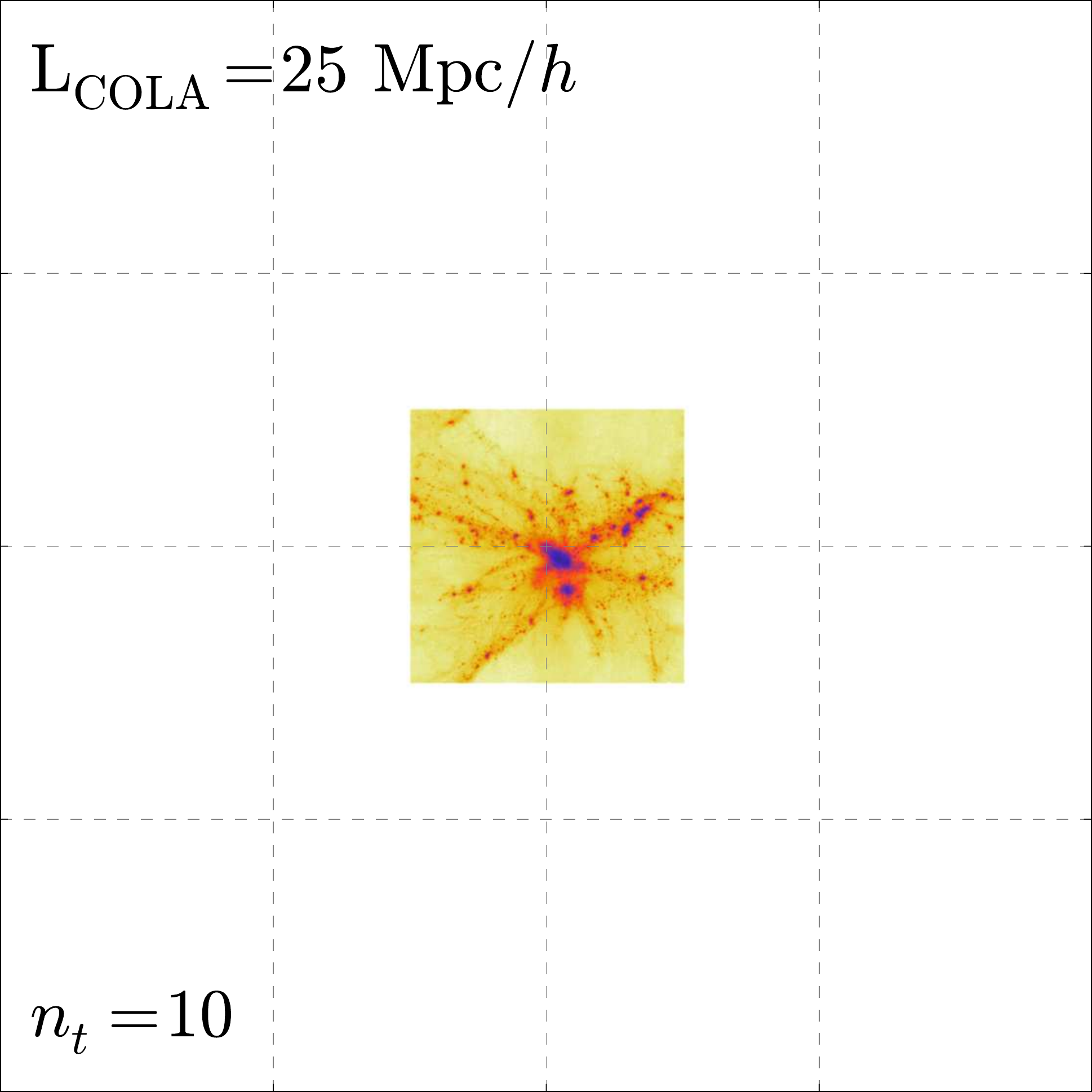}}
\hfill
\\
\vspace{-0.38cm}
\subfloat{\includegraphics[width=0.248\textwidth]{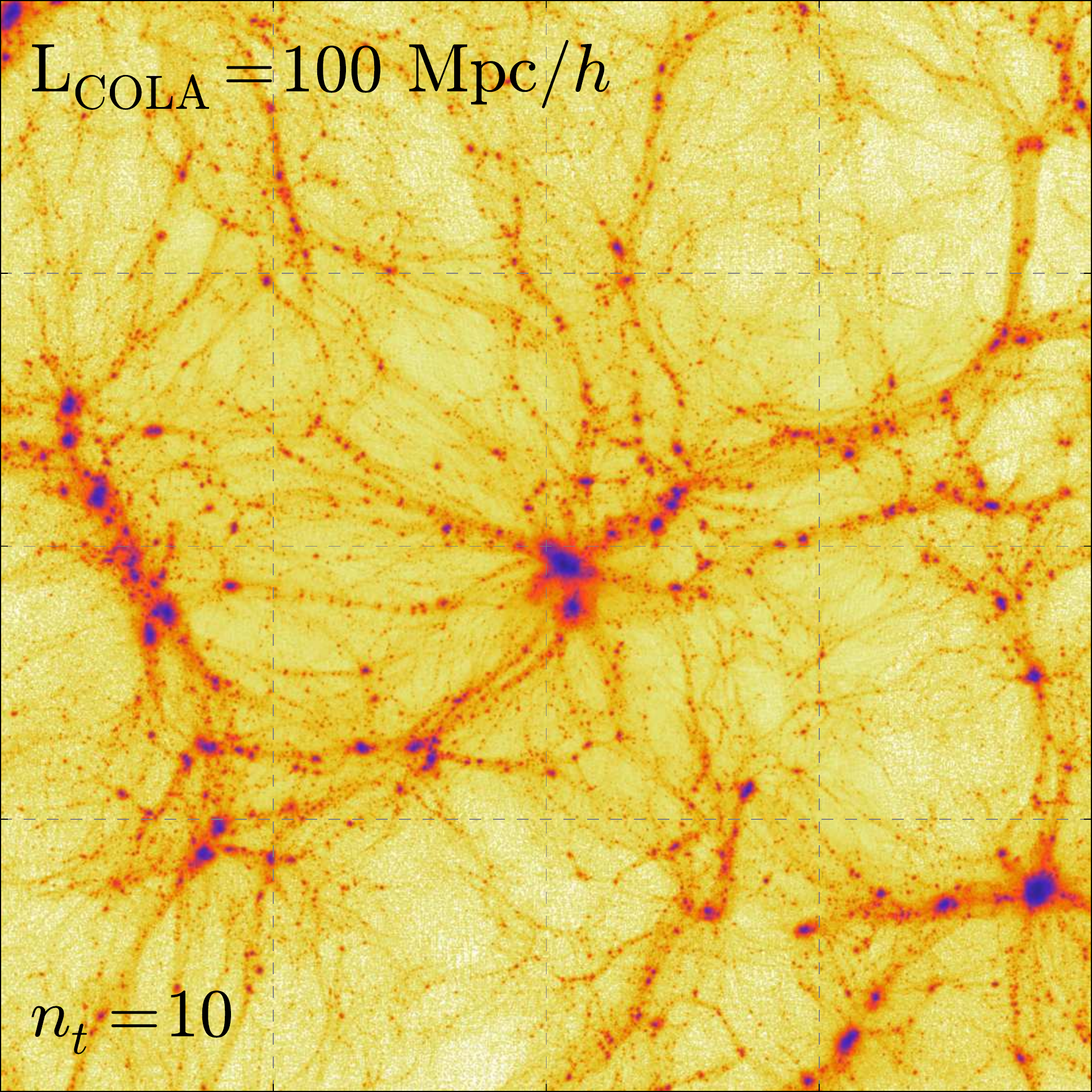}}
\subfloat{\includegraphics[width=0.248\textwidth]{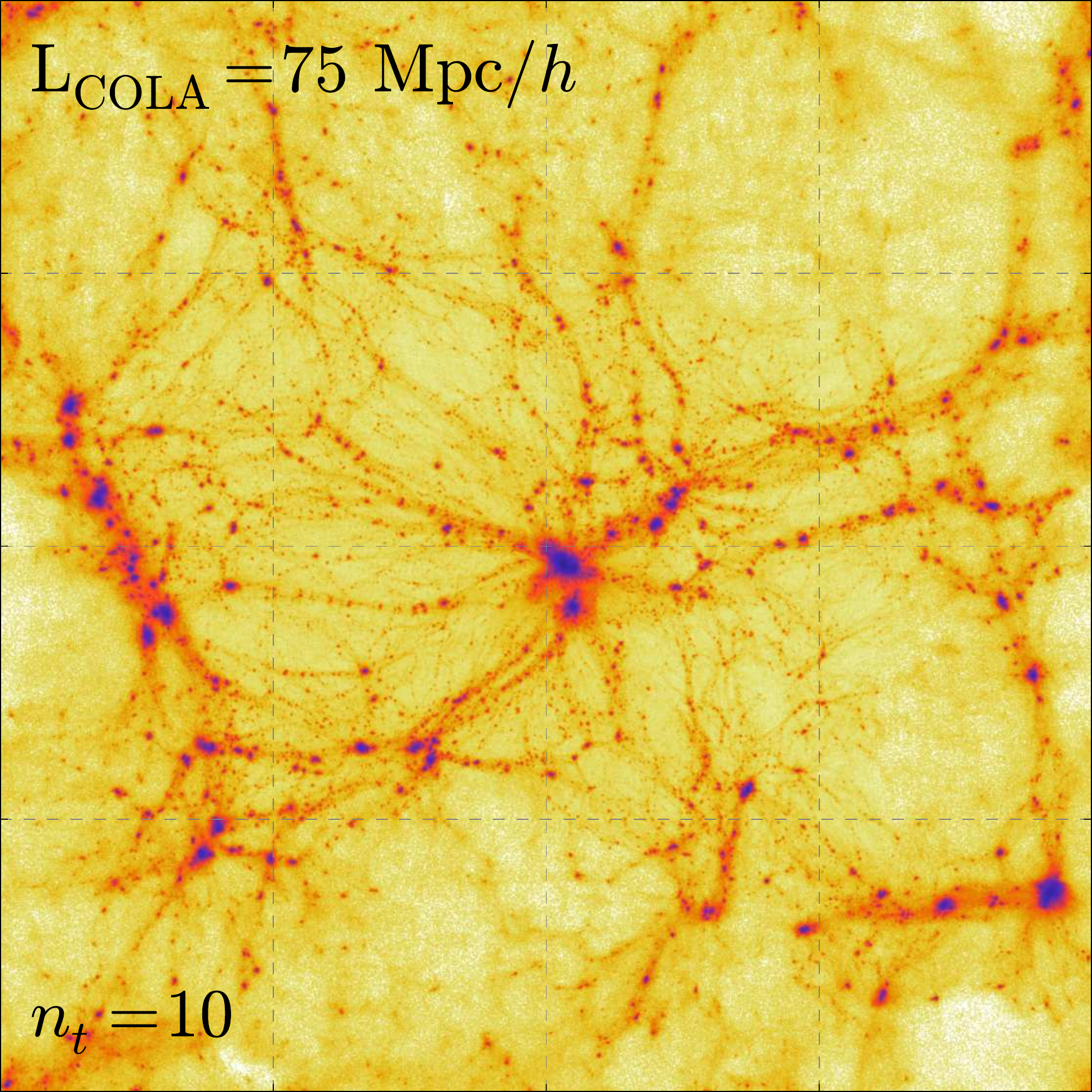}}
\subfloat{\includegraphics[width=0.248\textwidth]{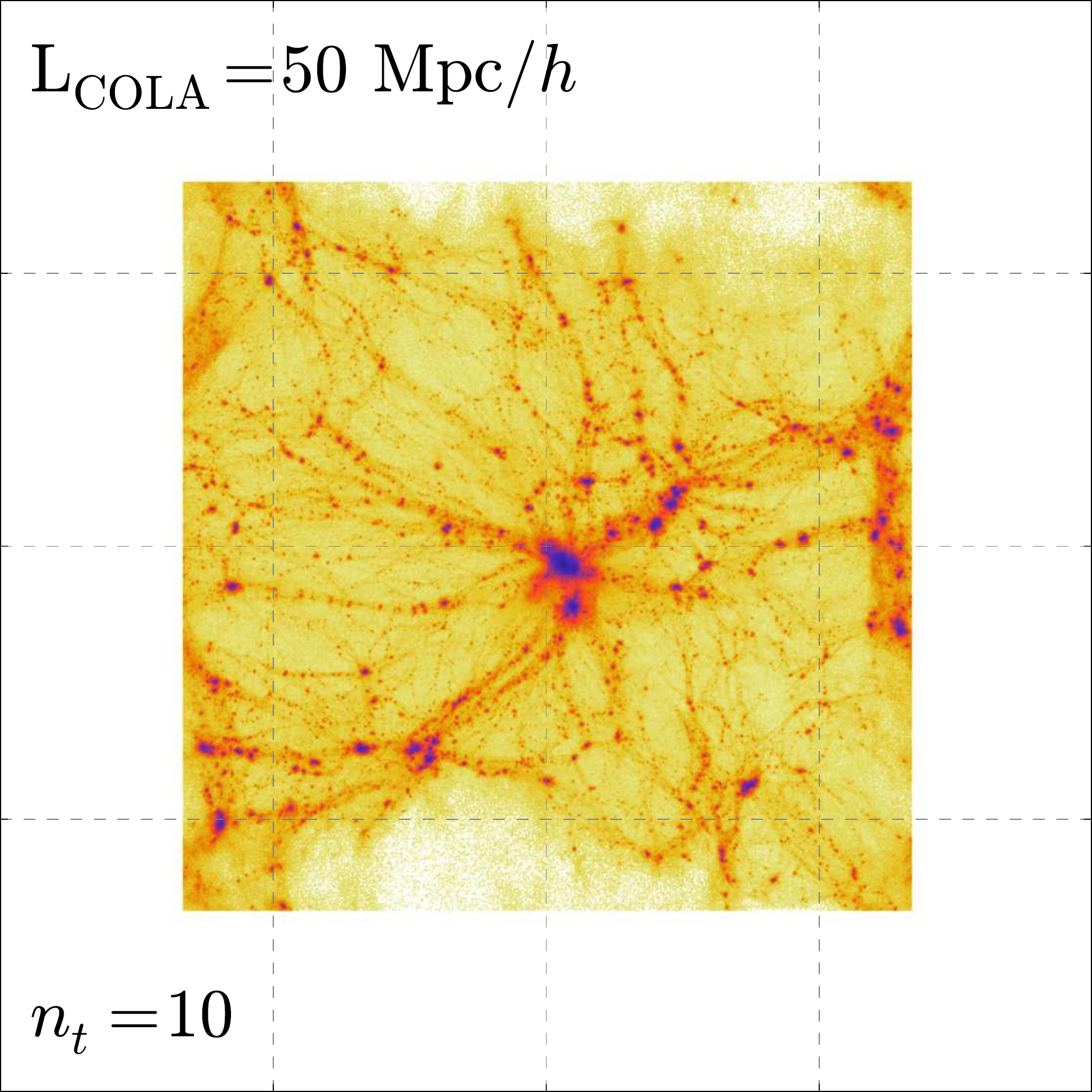}}
\subfloat{\includegraphics[width=0.248\textwidth]{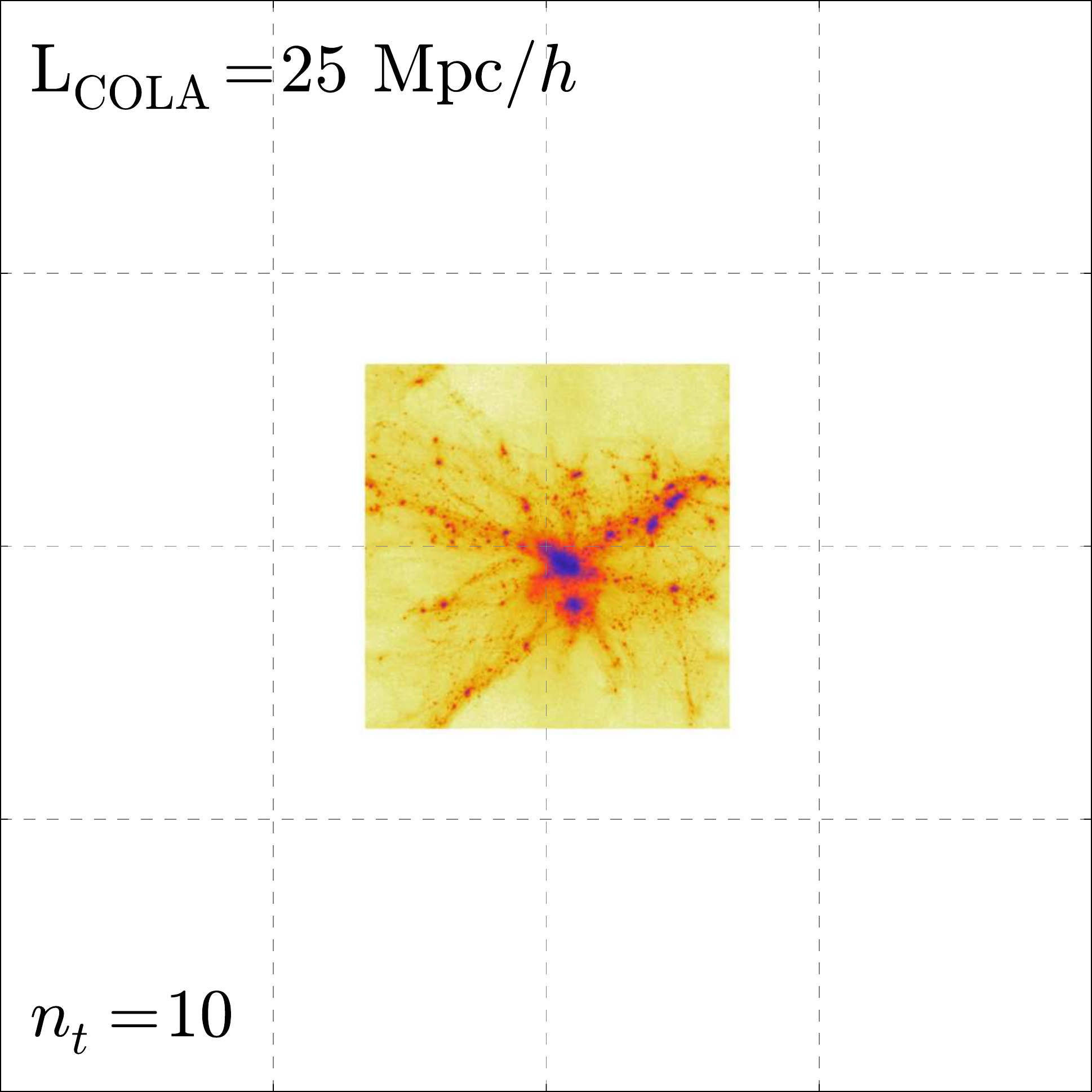}}
\hfill
\\
\vspace{-0.38cm}
\subfloat{\includegraphics[width=0.248\textwidth]{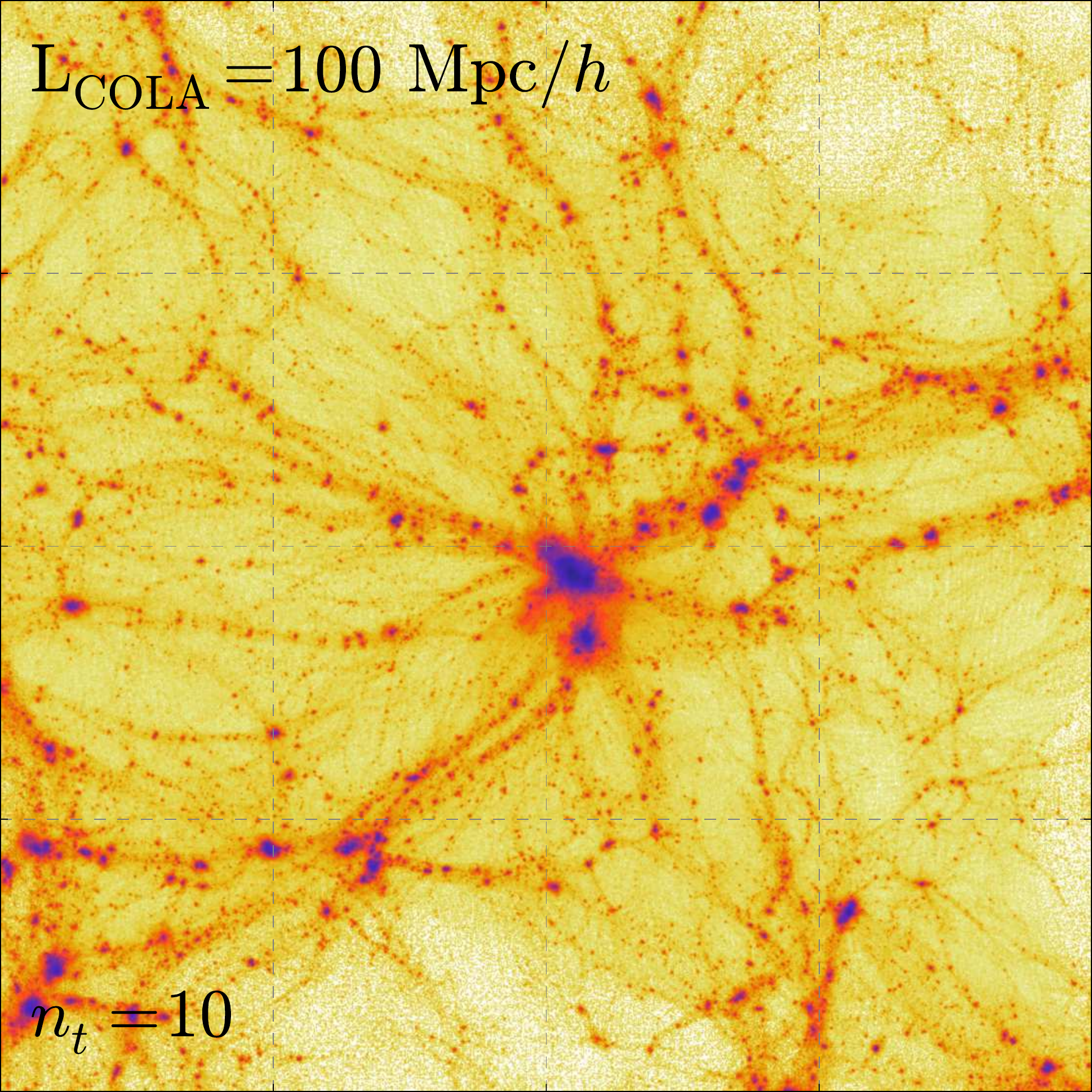}}
\subfloat{\includegraphics[width=0.248\textwidth]{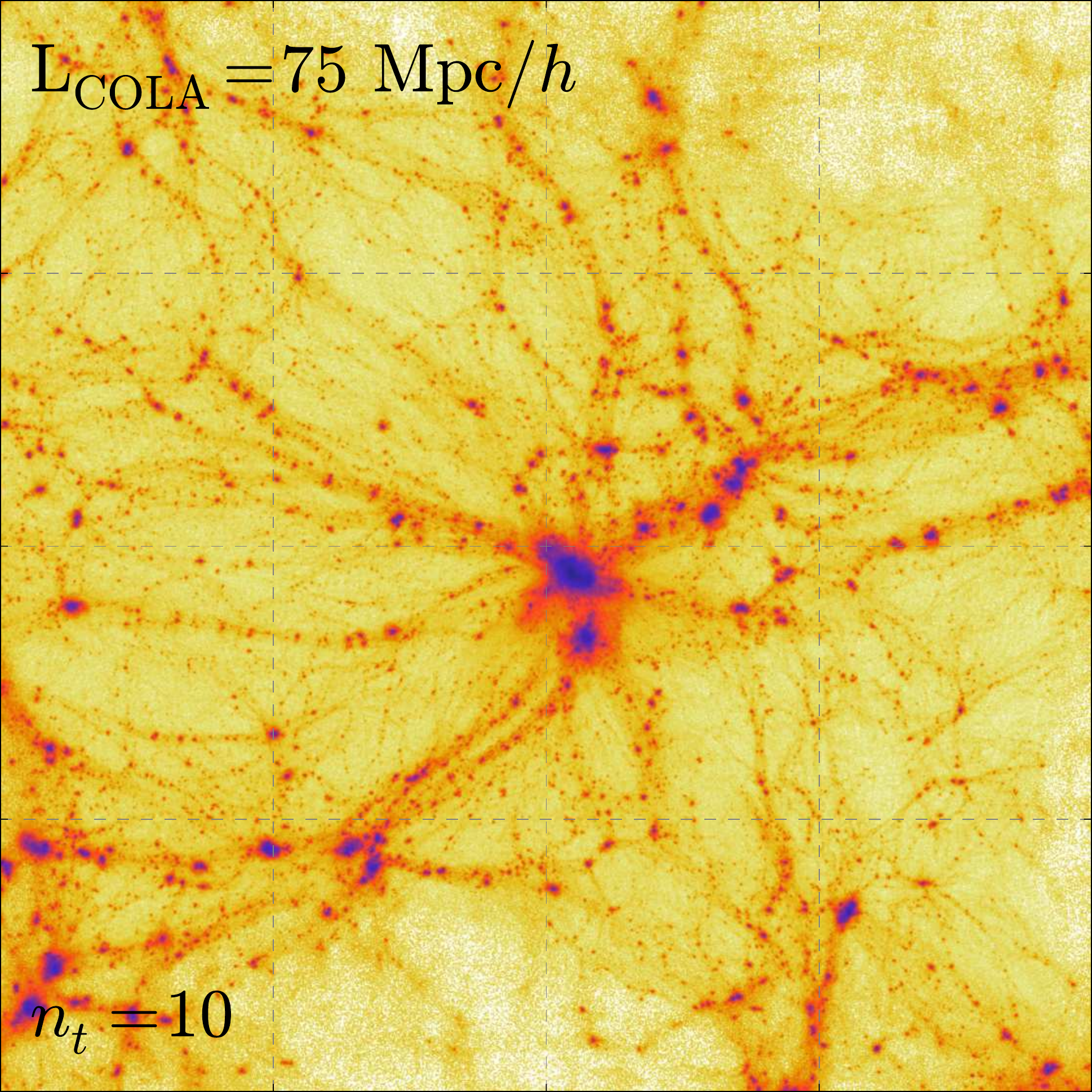}}
\subfloat{\includegraphics[width=0.248\textwidth]{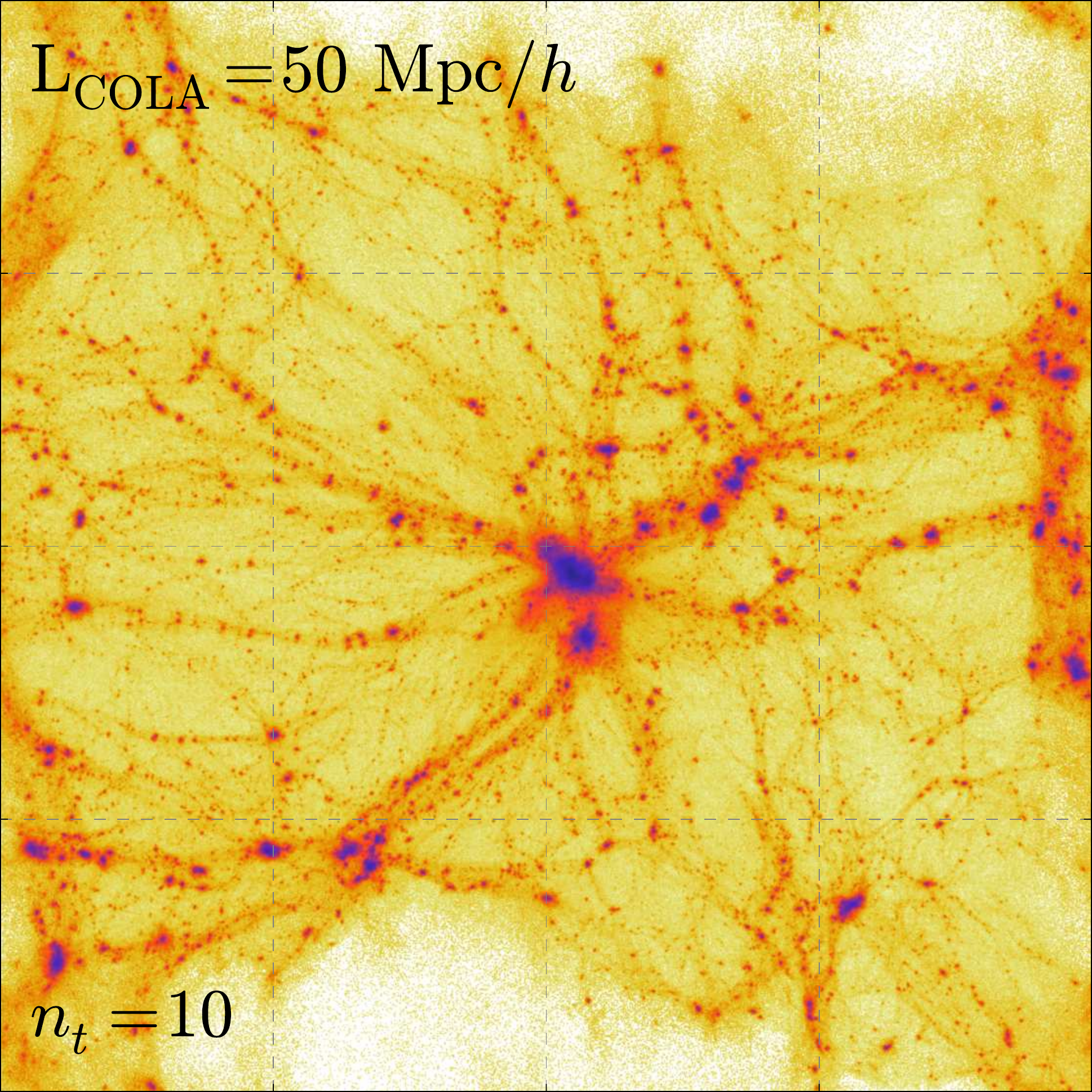}}
\subfloat{\includegraphics[width=0.248\textwidth]{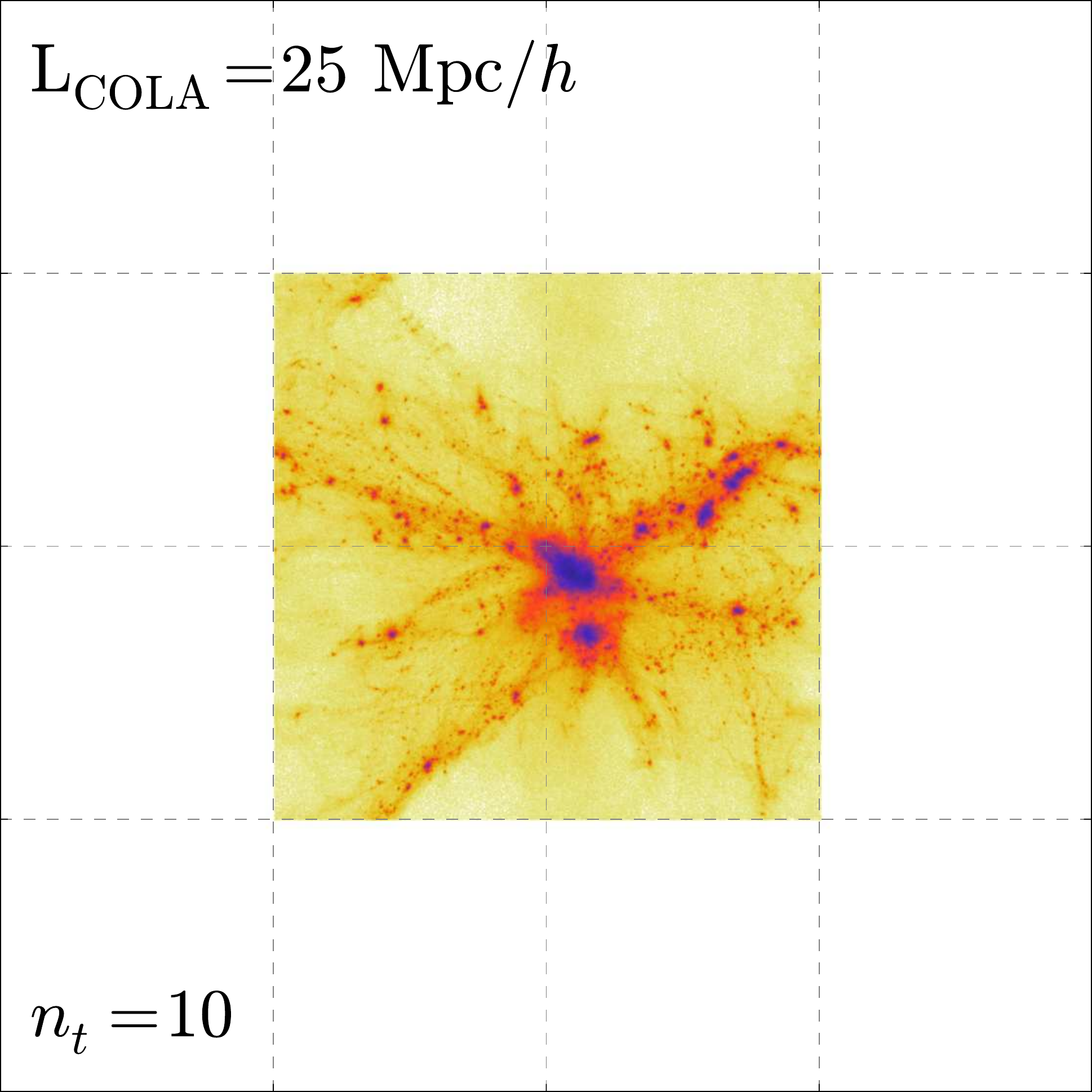}}
\hfill
\\
\vspace{-0.38cm}
\subfloat{\includegraphics[width=0.248\textwidth]{LAST_full_n10_cut0_L25.pdf}}
\subfloat{\includegraphics[width=0.248\textwidth]{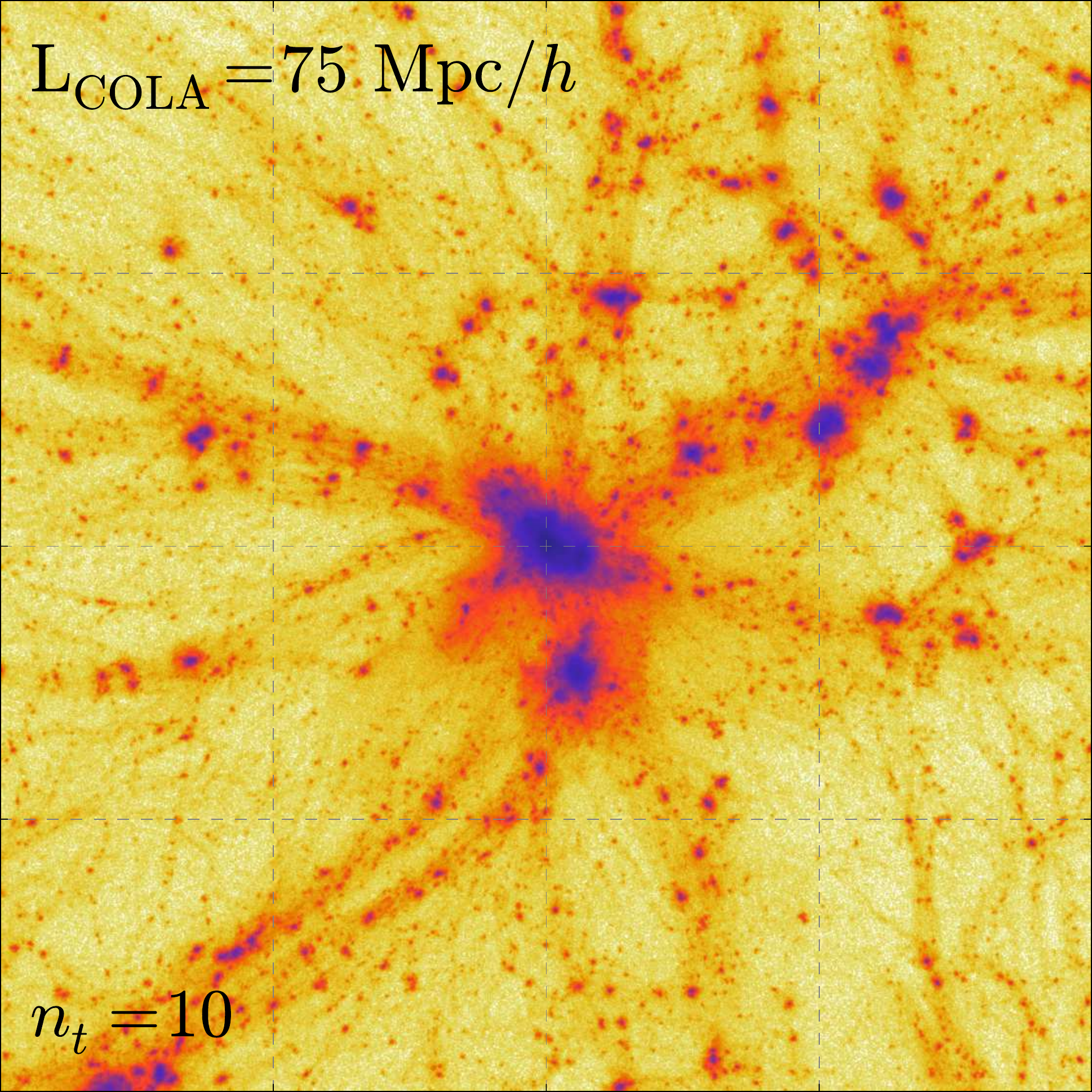}}
\subfloat{\includegraphics[width=0.248\textwidth]{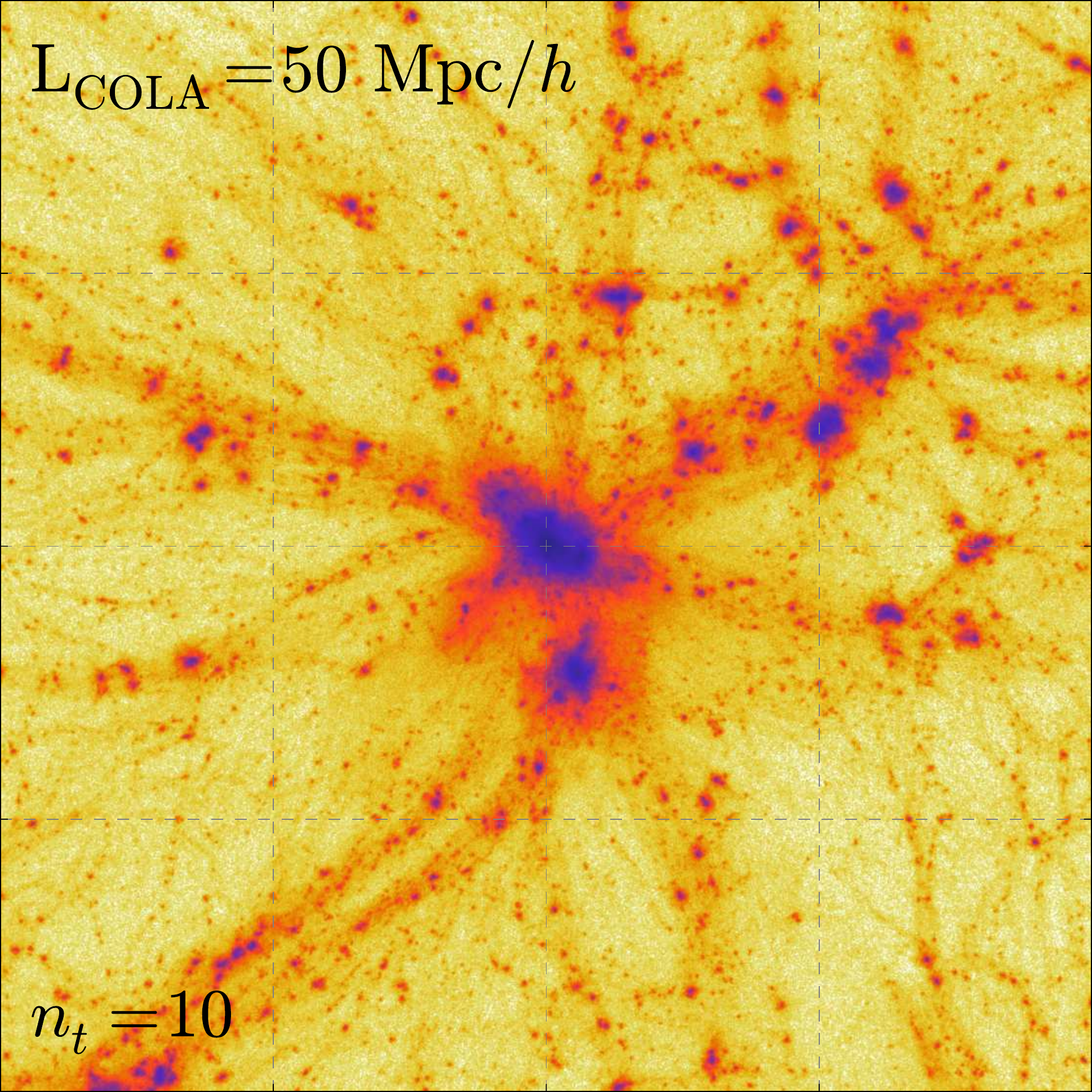}}
\subfloat{\includegraphics[width=0.248\textwidth]{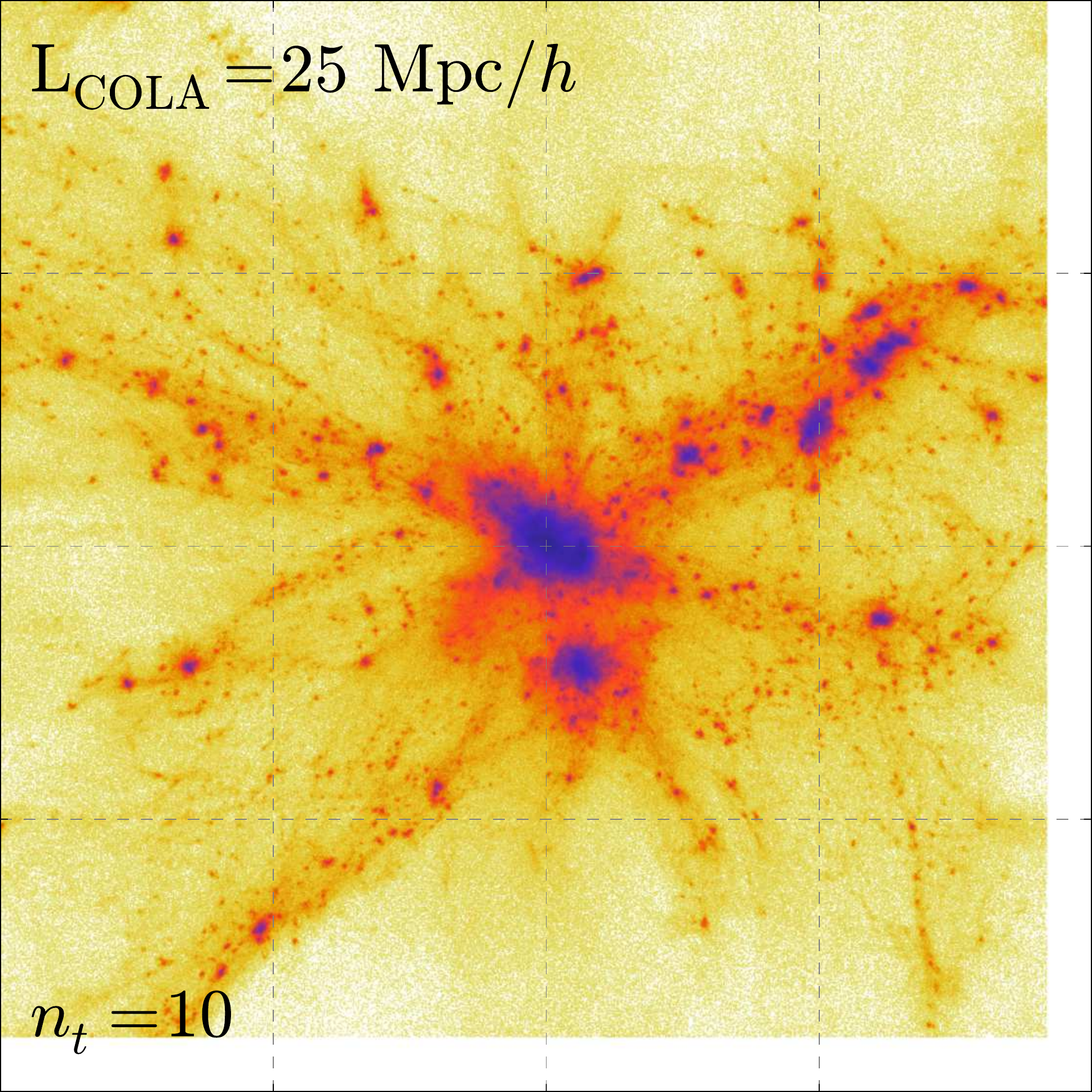}}
\hfill
\\
\vspace{-0.38cm}
\subfloat{\includegraphics[width=0.248\textwidth]{LAST_full_n10_cut0_L12_5.pdf}}
\subfloat{\includegraphics[width=0.248\textwidth]{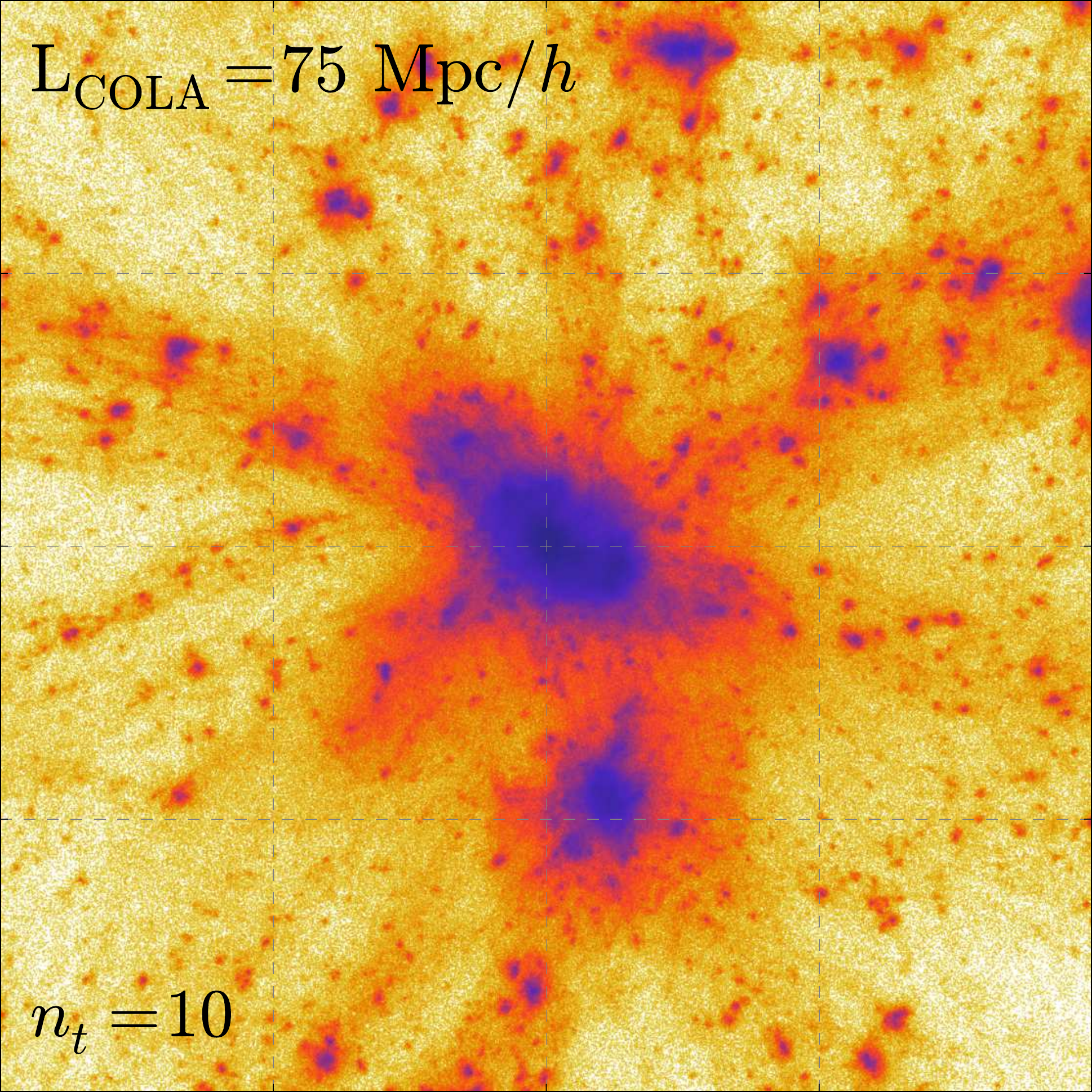}}
\subfloat{\includegraphics[width=0.248\textwidth]{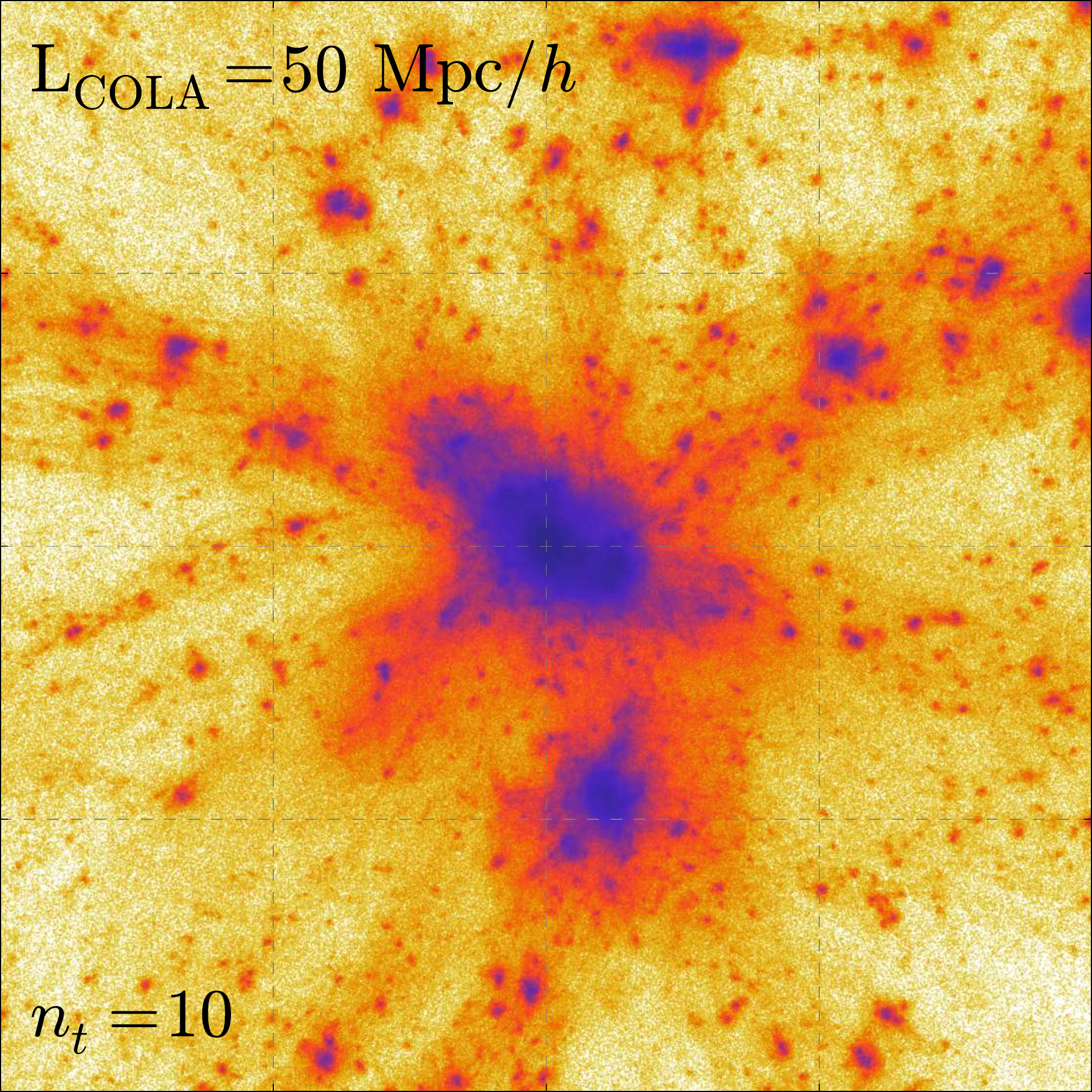}}
\subfloat{\includegraphics[width=0.248\textwidth]{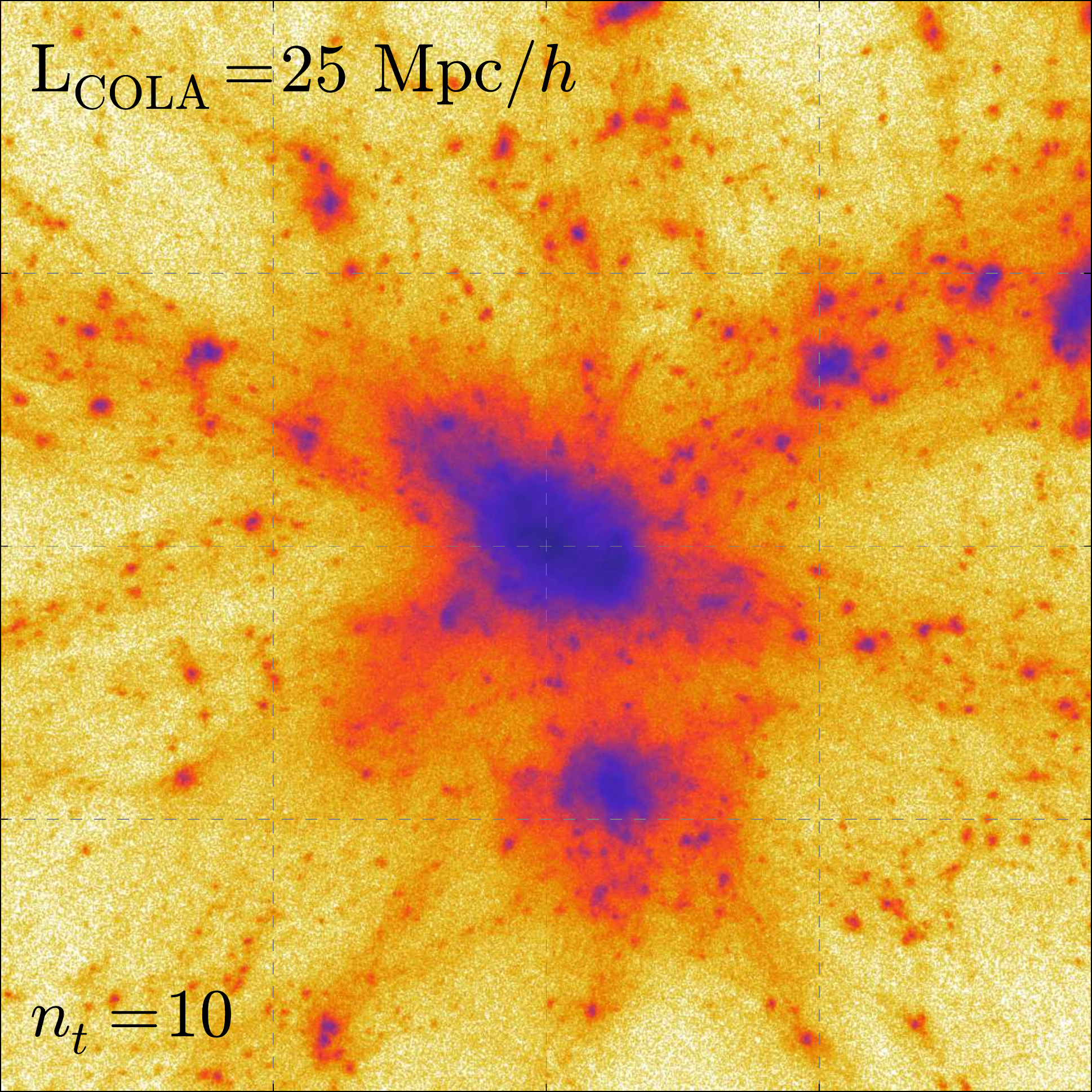}}
\hfill
\caption{The same as in Fig.~\ref{fig:slices100} but for $n_t=10$. This figure illustrates that COLA can be safely applied in the spatial and temporal domains simultaneously.} \label{fig:slices10}
\end{figure}

Next we focus on the simulation runs using the new sCOLA method. The results for $n_t=100$ are presented in Figure~\ref{fig:slices100}, while those for $n_t=10$ -- in  Figure~\ref{fig:slices10}. Columns represent different choices of COLA volume size, L$_{\mathrm{COLA}}$, while rows represent consecutive magnifications. Concentrating on the first row of each figure, one can see how the four different simulation volumes compare in size. Inspecting the last row, one can see that  the halo substructure is nearly identical across all panels with the exception of the rightmost panel (see below). Concentrating on the intermediate rows, we can see that large-scale structure is captured faithfully by sCOLA, and no large-scale distortions are introduced in the bulk even using clearly suboptimal \textit{periodic} boundary conditions for the COLA volumes.

The rightmost column (L$_{\mathrm{COLA}}=25$\,Mpc$/h$) shows clear problems: some substructure is missing, and the density field near the boundaries of the COLA volume is too smooth. The reason is straightforward to pinpoint and was already anticipated in Section~\ref{scola:disc}. The COLA box for that run is too small to include all particles that end up in the vicinity of the central halo, thus providing an illustration of how sCOLA behaves when it is pushed beyond breaking (see Section~\ref{scola:disc}).

It is not surprising that the snapshot with L$_{\mathrm{COLA}}=25$\,Mpc$/h$ exhibits problems. The boundary region, which is the region within $\sim10$\,Mpc$/h$ from the COLA boundary, where we expect to see sCOLA \hfill misbehaving \hfill  (see Section~\ref{scola:disc}), occupies most \hfill of the volume in this
\clearpage\noindent snapshot. Thus, it is only for L$_{\mathrm{COLA}}\sim50$\,Mpc$/h$ and above that we can expect the boundary effects on the central halo to be small, which is indeed confirmed.

Another example of sCOLA breaking is close to the COLA volume boundaries, which we again anticipated in Section~\ref{scola:disc}. In Figures~\ref{fig:slices100} and \ref{fig:slices10}, one can clearly see artifacts  from the non-zero $\bm{a}_{\mathrm{res}}$ and the periodic boundary conditions. Those are especially prominent on the right edge of the L$_{\mathrm{COLA}}=50$\,Mpc$/h$ COLA volume, where one can see a high-density wall of particles, which is missing from the larger COLA volume snapshots. Inspecting the larger COLA volume snapshots, we see that those particles are actually flowing to the left, outside of the central $50$\,Mpc$/h$, and are therefore wrapped around by the periodic BCs in the L$_{\mathrm{COLA}}=50$\,Mpc$/h$ snapshot. Similarly, the lower COLA boundary for the same snapshot is clearly emptied out since in our pyCOLA implementation no particles from outside the COLA volume are allowed to flow in.

Nevertheless, those boundary artifacts in the  L$_{\mathrm{COLA}}=50$\,Mpc$/h$ snapshot do not affect the accurate recovery of the density field in the bulk. Thus, in this example sCOLA has allowed us to reduce the simulation volume by a factor of 8 without sacrifice in the recovered central halo bulk properties, environment or substructures.

For $n_t=100$ the temporal domain behavior of sCOLA is approaching a full-blown N-body simulation. We next show how sCOLA behaves when $n_t$ is substantially reduced, thus making full use of the original temporal-domain COLA side of the method. Figure~\ref{fig:slices10}  demonstrates this feature of the method by significantly reducing both L$_{\mathrm{COLA}}$ and $n_t$. As one can see, those combined sCOLA/COLA snapshots share the same pros and cons of the two methods. Thus, depending on the application, one can apply COLA independently in the spatial and temporal domains, or simultaneously, as needed.

To be more quantitative, we investigate the
properties of the central halo.  We ran a friends-of-friends halo finding algorithm\footnote{We use the University of Washington FOF code available here: \url{http://www-hpcc.astro.washington.edu/tools/fof.html}} on all simulation snapshots presented in this paper, using a fixed linking length of 0.2. The results for the halo at the center of each snapshot are collected in Table~\ref{table:nt}. As reference values, we use the results from the
 most accurate simulation, corresponding to $L=$L$_\mathrm{COLA}=100$\,Mpc$/h$, $n_t=100$. Thus, the reference value for the mass of the central halo is $2.08\times 10^{14}$\,M$_\odot/h$.  In the table we list the error in the recovered halo mass relative to the reference value, the error in its center-of-mass position and velocity\footnote{For velocity, we quote values for $(d\x/d\eta)/(aH(a))$, where $H(a)$ is the Hubble parameter at scale factor $a$, $\eta$ is conformal time, and $\x$ is comoving distance. Thus, the units of our velocity variable are Mpc$/h$.  This definition allows calculating redshift-space positions trivially: one simply has to
add the line-of-sight velocity to the halo position.}  and the error in the root-mean-square (rms) of the halo particle positions, giving a measure of the halo size, as well as the error in the rms halo particle velocities, giving a measure of the FoG effect in redshift space.

In Table~\ref{table:nt}, we can see that for $n_t=100$, sCOLA performs superbly for L$_{\mathrm{COLA}}=75$\,Mpc$/h$ and  L$_{\mathrm{COLA}}=50$\,Mpc$/h$, i.e. as long as all halo particles are captured inside the COLA volume. All errors in both real and redshift space are around or well within the force resolution, and the halo mass is recovered to 1\% or better. 

When $n_t$ is pushed down to $10$, we see that most of the deterioration can be attributed to the smaller number of timesteps (i.e. to the original COLA method), and not to the sCOLA side of the code. That is evident when comparing the results of smaller COLA volumes to the result of the  L$_{\mathrm{COLA}}=100$\,Mpc$/h$, $n_t=10$ snapshot.

\section{Summary}\label{sec:sum}

In this proof-of-concept paper we  presented sCOLA -- an extension to the N-body COLA method in the spatial domain. sCOLA is a straightforward modification to N-body codes, which uses perturbation theory to calculate the far field potential from masses outside a region of interest, while solving for the near field using an N-body code. The use of perturbation theory allows for the N-body side of sCOLA to perform all calculations entirely within the region of interest (dubbed the ``COLA volume''), without the need to substantially increase the simulated volume to capture the effects of large modes or structures outside that region. 

Compared to standard zoom-in simulation methods, the main advantage of sCOLA is that it makes the different volumes \textit{completely} independent once the initial conditions are computed.  Thus, there is no need to add lower resolution zones around the region of interest in sCOLA. 

The sCOLA method leads to the exact equations of motion when the COLA volume coincides with the full volume, capturing all large-scale modes affecting the region of interest. When the COLA volume does not cover the full volume, the approximations behind sCOLA require one to run the N-body code behind sCOLA on a region which is roughly 10\,Mpc$/h$ larger in all directions than the region of interest.

We demonstrated the usage of sCOLA by showing results from zoom-in simulations performed with a freely available sCOLA-based N-body code (called pyCOLA), which we developed for this paper. Our results demonstrate that sCOLA faithfully recovers large-scale structure inside a COLA volume as small as 50\,Mpc$/h$ (possibly even smaller), while keeping substructures intact. Moreover, we showed that sCOLA can be safely used in conjunction with the original COLA method. We find that reducing the number of timesteps for COLA naturally degrades the quality of the simulation. However, the sCOLA-side of the method proves robust to reducing the size of the COLA box until the COLA boundary region (at redshift zero, corresponding to the region $\sim10$\,Mpc$/h$ or less from the edge) invades the region of interest.

An immediate application of sCOLA is as a method for performing zoom-in simulations, which we illustrated in this paper. The fact that the N-body side of sCOLA is needed to perform calculations of the potential only for the near field, implies that sCOLA-based N-body codes can be made embarrassingly parallel, thus opening the door for efficiently tiling a volume of interest using grid computing in what may become Cosmology@Home. Moreover, combining sCOLA with the original COLA method can be useful for cheaply generating large ensembles of accurate mock halo catalogs required to study uncertainties of large-scale structure observables. Surveys with large aspect ratios, such as pencil-beam surveys, can especially benefit, as sCOLA can easily capture the effects of large-scale transverse modes without the need to substantially increase the simulated volume.

pyCOLA can be found on the following URL: \url{https://bitbucket.org/tassev/pycola/}.

\acknowledgments
ST would like to thank his colleagues and students at Braintree High School for making his entrance into public high school teaching a rewarding and joyful experience. M.Z. is supported in part by NSF Grants No. PHY-1213563 and No. AST-1409709. 

\appendix

\section{The details behind sCOLA}\label{app:KDK}

In this appendix we write down the exact first equation of the sCOLA algorithm, given schematically by eq.~(\ref{sCOLA1}). One should read this appendix as a continuation of Appendix A of TZE. The preliminaries, notation, and original temporal-domain COLA (tCOLA) are presented there. Here we present  the modifications to the original tCOLA method needed to extend COLA to the spatial domain.

\subsection{The sCOLA method}\label{app:KDKALC}

The COLA method (both tCOLA and sCOLA) dictates that we integrate the equation of motion for CDM particles in a frame comoving with observers following trajectories  specified by LPT. Thus, we solve for the residual displacement of CDM particles:
\be
\bm{s}_\mathrm{res}\equiv \bm{s}-D_1\bm{s}_1-D_2\bm{s}_2\ .
\ee
For tCOLA, the equation of motion is written as (see eq. A.6 in TZE)
\be\label{eomALC}
T^2[\bm{s}_\mathrm{res}]=-\frac{3}{2}\Omega_M a\partial_{\x}\partial_{\x}^{-2}\delta(\x,a)-T^2[D_1]\s_1-T^2[D_2]\s_2\ .
\ee
For sCOLA, we rewrite it as
\be\label{eomALCs}
T^2[\bm{s}_\mathrm{res}]=-\frac{3}{2}\Omega_M a\partial_{\x}\partial^{\bx,-2}_{\x}\delta(\x,a)-T^2[D_1]\s^{\bx}_1-T^2[D_2]\s^{\bx}_2\ .
\ee
Thus, the only difference between sCOLA and tCOLA is adding the \textbf{COLA} superscript where appropriate, reminding us that those quantities are calculated in the small COLA box, covering only a fraction of the original volume.  The calculation of the LPT displacement fields, $\s^{\bx}_{1,2}$, in the COLA box is presented in Appendix~\ref{app:KDKlpt}.

Below we present results for the combined tCOLA/sCOLA discretization as employed in pyCOLA.
Therefore, as described in TZE, next we discretize the operator $T$ only on the left-hand side of the above equation, using as a velocity variable $\v_{\mathrm{res}}\equiv T[\s_{\mathrm{res}}]$, which is the CDM velocity as measured by an LPT observer.

\subsection{Discretizing $T$ in sCOLA in the standard approach}

Similar to TZE, we discretize the sCOLA equations using a kick-drift-kick scheme. That requires writing down a kick operator $\mathrm{K}$, a drift operator $\mathrm{D}$, as well as introducing two more operators, $\mathrm{L}_\pm$, for converting between the standard particle velocity  and the residual velocity as measured by LPT observers. 

In the ``standard'' prescription (see A.3.1 of TZE), the operators $\mathrm{D}$ and $\mathrm{L}_\pm$ of tCOLA and sCOLA are identical. The kick operator in sCOLA, however, is modified (compare with eq. A.7 of TZE):
\be\label{KDKALCstd}
\mathrm{D}(a_i,a_f;a_c):& \ \  &\x(a_i) \ \mapsto 
\  \x(a_f)=\x(a_i)+\v(a_c) \int\limits_{a_i}^{a_f} \frac{d\tilde a}{Q(\tilde a)}+
\\
&&+(D_1(a_f)-D_1(a_i))\s_1+(D_2(a_f)-D_2(a_i))\s_2\ ,\nonumber
\\
\mathrm{K}(a_i,a_f;a_c):&  \ \ &\v(a_i) \ \mapsto  
\ 
\v(a_f)=\v(a_i)-\left[\int\limits_{a_i}^{a_f} \frac{\tilde a/a_c}{Q(\tilde a)}d\tilde a\right]\times\nonumber\\
&&
\left(
-\frac{3}{2}\Omega_M a_c \partial_{\x}\partial_{\x}^{\bx,-2} \delta(\x,a_c)-T^2[D_1](a_c)\s^{\bx}_1-T^2[D_2](a_c)\s^{\bx}_2
\right)\ ,\nonumber
\ee
where as in TZE we dropped the $(\mathrm{res})$ subscript from $\v_\mathrm{res}$ for brevity.

Time evolution between $a_0$ and $a_{n+1}$ in sCOLA is achieved in exactly the same way as in tCOLA, by applying the following operator on $(\x(a_0), \v(a_0)=T[\s](a_0))$:
\be\label{ALCtimestep}
\mathrm{L}_+(a_{n+1})\left(\prod_{i=0}^n\mathrm{K}(a_{i+1/2},a_{i+1};a_{i+1})\mathrm{D}(a_i,a_{i+1};a_{i+1/2})\mathrm{K}(a_i,a_{i+1/2};a_i)\right)\mathrm{L_-}(a_{0})\ .
\ee
As in TZE, here we defined 
\be\label{Lops}
\mathrm{L}_\pm(a):\ \  \v(a) \ &\mapsto&\  \v(a)=\v(a)\pm\bigg(T[D_1](a) \s_1+T[D_2](a) \s_2\bigg)\ ,
\ee
which as mentioned above, first transforms   the initial conditions for $\v$ to the rest frame of LPT observers ($L_-$), and then adds back the LPT velocities at the end ($L_+$).

\subsection{Modified discretization of $T$ in sCOLA}\label{app:modKDK}

As done in Appendix A.3.2 of TZE, here we present a modified discretization scheme of $T$, which allows one to improve the convergence properties of COLA. As above, the $\mathrm{L}_\pm$ and $\mathrm{D}$ operators of tCOLA and sCOLA in this scheme are identical. The kick operator in sCOLA, however, is modified (compare with eq. A.15 of TZE):
\be\label{KDKALC}
\mathrm{D}(a_i,a_f;a_c):& \ \  &\x(a_i) \ \mapsto\  \x(a_f)=\x(a_i)+\frac{\v(a_c)}{u(a_c)} \int\limits_{a_i}^{a_f} \frac{u(\tilde a)}{Q(\tilde a)}d\tilde a+\\
&&+(D_1(a_f)-D_1(a_i))\s_1+(D_2(a_f)-D_2(a_i))\s_2\ ,\nonumber
\\
\mathrm{K}(a_i,a_f;a_c):& \ \ &\v(a_i) \  \mapsto  
\v(a_f)=\v(a_i)+\frac{u(a_f)-u(a_i)}{T[u](a_c)}\times\nonumber\\
&&
\left(
-\frac{3}{2}\Omega_M a_c \partial_{\x}\partial_{\x}^{\bx,-2} \delta(\x,a_c)-T^2[D_1](a_c)\s^{\bx}_1-T^2[D_2](a_c)\s^{\bx}_2
\right)\ ,\nonumber
\ee
where we again dropped the $(\mathrm{res})$ subscript from $\v_\mathrm{res}$ for brevity.
The equations presented above give the evolution equations behind the sCOLA algorithm presented in this paper.

However, in practice, we found that using the above prescription results in density fields from sCOLA which are artificially smoothed. The reason is due to a mismatch at very short scales between the LPT displacement fields calculated in the sCOLA box and the original simulation volume. Such a mismatch at deeply non-linear scales has no deep physical origin and we view it as an artifact of the different algorithms we use for calculating the LPT fields, combined with numerical precision issues. Removing that artifact is trivial: we boxcar average\footnote{For different simulation set-ups, one may need to experiment with varying the boxcar window, or using a Gaussian averaging instead.} (with width of 2 particle cells) all LPT displacement fields entering the operators $\mathrm{L}_\pm$, $\mathrm{K}$ and $\mathrm{D}$. That eliminates most short-scale noise, which otherwise artificially smoothed our final results. We would like to emphasize that no information is lost in such an averaging as LPT carries little (if any) information on the scale of the averaging window we chose. Moreover, the convergence of our scheme to the ``truth'' when the number of timesteps and sCOLA volume are brought to infinity is not affected since in that case we would add and subtract the same averaged LPT displacements. Note that the averaging is done only on the displacements entering the evolution operators, and \textit{not} on the initial conditions, as we want to keep the initial conditions intact. 

\section{Calculating LPT displacements using force evaluations}\label{app:KDKlpt}

In this appendix we write down the exact second equation of the sCOLA algorithm, given schematically by eq.~(\ref{sCOLA2}), and describe how we solve it in pyCOLA. The non-schematic version of equation (\ref{sCOLA2}) reads:
\be\label{sCOLAext}
T^2[D_1]\s^{\bx}_{1}+T^2[D_2]\s^{\bx}_{2}= -\frac{3}{2}\Omega_M a\bm{\nabla} \nabla^{-2}_{\bx}\delta[\x_{\mathrm{LPT}}](\x_{\mathrm{LPT}}) \ \ \  \hbox{(solved in PT),}\nonumber\\
\ee
where $\x_{\mathrm{LPT}}=\q+D_1\bm{s}_1+D_2\bm{s}_2$. 

The usual way to derive the LPT displacement fields, $\s_1$ and $\s_2$, is using  Fourier techniques \cite{2006MNRAS.373..369C}.  However, the method we employ is based on \cite{garrison}, where the equation above is solved in real space by calculating the gravitational forces due to the density field entering above at early enough times that one can consider the results perturbative. The right-hand side above is completely determined by the initial conditions in the full volume. So, all we need to do is extract its first- and second-order pieces and match those to the corresponding terms on the left-hand side. We do that explicitly below, following \cite{garrison}. This is advantageous in our case because the displacement fields in the equation above live in two separate boxes -- the COLA box ($\s^{\bx}_{1,2}$) and the full box ($\s_{1,2}$). The results are summarized below, and are  implemented in the public pyCOLA code. These results include modifications to \cite{garrison} (allowing $\beta=1$ in the notation below) to obtain high-precision LPT displacements for the sCOLA box. For further information, see \cite{garrison}.

    pyCOLA offers a choice in the calculation of the first-order and second-order displacements, $\bm{s}_{1,2}^{\bx}$, in the COLA volume, as well as $\bm{s}_{2}$ in the full volume at 
    redshift zero: using a 2-pt or  a
    4-pt (denoted by superscript) rule. 
Depending on the rule, the first/second order 
       displacements receive higher-order corrections as follows: at third/fourth order when using the 2pt rule; at fifth/sixth 
       order when using the 4pt rule.  The displacement fields are given by ($\eta$ below stands for the displacement field in the $\bx$ box or the full box):
      \begin{eqnarray}\label{forcess}
          \bm{s}^{\eta,\mathrm{2pt}}_{1}    & = & - \frac{1}{2g}                      \left[\bm{F}_{\eta}(g,\beta g^2)-\bm{F}_{\eta}(-g,\beta g^2)\right] \ ,\nonumber\\
          \bm{s}^{\eta,\mathrm{2pt}}_{2}    & = & - \frac{\alpha}{2g^2}                 \left[\bm{F}_{\eta}(g,\beta g^2)+\bm{F}_{\eta}(-g,\beta g^2)\right] \ ,\nonumber\\
          \bm{s}^{\eta,\mathrm{4pt}}_{1}    & = & - \frac{1}{2g}     \frac{a^2}{a^2-1}\bigg[\bm{F}_{\eta}(g,\beta g^2)-\bm{F}_{\eta}(-g,\beta g^2)- \nonumber\\
                                                     &   & \quad \quad \quad \quad \quad \quad - \frac{1}{a^3}\bigg(\bm{F}_{\eta}\left(a g,\beta a^2 g^2\right)-\bm{F}_{\eta}\left(-a g,\beta a^2 g^2\right)\bigg)\bigg] \ ,\nonumber\\
          \bm{s}^{\eta,\mathrm{4pt}}_{2}    & = & - \frac{\alpha}{2g^2}\frac{a^2}{a^2-1}\bigg[\bm{F}_{\eta}(g,\beta g^2)+\bm{F}_{\eta}(-g,\beta g^2)-\nonumber\\
                                                     &   & \quad \quad \quad \quad \quad \quad - \frac{1}{a^4}\bigg(\bm{F}_{\eta}\left(a g,\beta a^2 g^2\right)+\bm{F}_{\eta}\left(-a g,\beta a^2 g^2\right)\bigg)\bigg] \ ,
      \end{eqnarray}
where
\be
\alpha&=&(3/10)\Omega_{M}^{1/143}  \ \hbox{if }\ \beta=1\ ,\nonumber\\
\alpha&=&(3/7)\Omega_{M}^{1/143} \  \, \ \hbox{if }\ \beta=0\ ,
\ee
and $a\neq 1$ and $g$ are arbitrary constants. The equations above become exact in the limit of $g\to 0$,  $ag\to 0$. 
When writing the equations above in the COLA box to second order (i.e. $\eta=\bx$), one must use $\beta=1$ in order to be consistent in PT. If one wants to use the above equations for  calculating $\bm{s}_{2}$ in the full box, then one must set $\beta=0$ since $\beta=1$ requires one to provide $\bm{s}_{2}$. The factors of $\Omega_{M}^{1/143}$ ($\Omega_M$ being 
    the matter density today) are needed to rescale the second order 
    displacements to matter domination and are correct to 
    $\mathcal{O}(\max(10^{-4},g^3/143))$ in 
    $\Lambda\mathrm{CDM}$. 
    
    The force $\bm{F}_\eta(g_1,g_2)$ corresponds to the right-hand side of eq.~(\ref{sCOLAext}). It is evaluated numerically by using the force calculation function (using a 4-pt finite difference scheme) of pyCOLA on the PM grid. Analytically, $\bm{F}_\eta$ is given by:
      \begin{eqnarray}
          \bm{F}_\eta(g_1,g_2) = \bm{\nabla}\nabla_\eta^{-2}\delta\left[\q+g_1\bm{s}_1+g_2\Omega_{M}^{-1/143}\bm{s}_2\right]\ ,
      \end{eqnarray}
where $g_1$ and $g_2$ correspond to the first and second order growth factors, assumed to be deep in matter domination (see the arguments of $\bm{F}$ in eq.~(\ref{forcess}) above). 

The fractional overdensity, $\delta$, in pyCOLA is calculated using cloud-in-cell assignment from a grid of particles (at Lagrangian positions $\q$) displaced by the first and second order displacement  fields: $g_1\bm{s}_1+g_2\Omega_{M}^{-1/143}\bm{s}_2$. Note that implicitly for each particle at 
    Lagrangian position $\bm{q}$, the force 
    $\bm{F}_\eta(g_1,g_2)$ is evaluated at its corresponding Eulerian position:
    $\bm{q}+g_1\bm{s}_1+g_2\Omega_{M}^{-1/143}\bm{s}_2$. The choice of $\eta$ ($\eta=\bx$ or $\eta$=``full'') determines which particles (only those in the COLA volume, or all particles in the full volume) are included in the calculation of the potential, $\nabla_\eta^{-2}\delta$.

\bibliography{mildly_NL_v2}

\end{document}